\documentclass[twocolumn, 9pt]{aastex62}
\usepackage{wrapfig} 
\usepackage{savesym}
\savesymbol{tablenum} 
\usepackage{siunitx}
\usepackage{multirow}

\def\SBunits{erg s$^{-1}$ cm$^{-2}$ arcsec$^{-1}$}
\def\fluxunits{erg s$^{-1}$ cm$^{-2}$}
\def\lumunits{erg s$^{-1}$}

\def\noobjs{five} 
\def\nohalos{four} 

\def\Roche{\mbox{J2228$+$0110}} 
\def\Comp{\mbox{J2100$-$1715}}
\def\Italian{\mbox{J1030$+$0524}}
\def\CFHQS{\mbox{J2329$-$0301}} 
\def\Ffav{\mbox{P231$-$20}}

\def\RocheTotFlux{\num{4.64E-16}}
\def\RocheTotLum{44.250}
\def\ItalianTotFlux{\num{4.28E-17}}
\def\ItalianTotLum{43.282} 
\def\CFHQSTotFlux{\num{1.49E-16}}
\def\CFHQSTotLum{43.841}
\def\FfavTotFlux{\num{1.77E-16}}
\def\FfavTotLum{43.942}

\def\RocheRadArcsec{$5.1$ arcsec}
\def\RocheRadPkpc{$29.1$ pkpc}
\def\ItalianRadArcsec{$2.8$ arcsec}
\def\ItalianRadPkpc{$15.7$ pkpc} 
\def\CFHQSRadArcsec{$2.2$ arcsec}
\def\CFHQSRadPkpc{$12.2$ pkpc}
\def\FfavRadArcsec{$3.4$ arcsec}
\def\FfavRadPkpc{$18.1$ pkpc}

\def\SBlimNBRoche{\num{6.43E-18}} 
\def\SBlimNBComp{\num{4.84E-18}} 
\def\SBlimNBItalian{\num{3.23E-18}}
\def\SBlimNBCFHQS{\num{6.80E-18}}
\def\SBlimNBFfav{\num{7.11E-18}}

\def\layersNBRoche{$64$}
\def\layersNBComp{$40$}
\def\layersNBItalian{$28$}
\def\layersNBCFHQS{$36$}
\def\layersNBFfav{$32$}

\def\SBlimLyaRoche{\num{5.03E-19}} 
\def\SBlimLyaComp{\num{6.45E-19}} 
\def\SBlimLyaItalian{\num{4.71E-19}} 
\def\SBlimLyaCFHQS{\num{7.40E-19}}
\def\SBlimLyaFfav{\num{8.81E-19}}

\def\GaussianSmoothingKernelSize{1.2 pixels} 
\def\contourhigh{\num{1.0E-17}}

\def\gtSBbright{SB $> 10^{-17}$ erg s$^{-1}$ cm$^{-2}$ arcsec$^{-2}$}

\def\avgOffsetArcsecs{$0.77$ arcsecs}
\def\avgOffsetpkpc{$4.3$ pkpc}

\graphicspath{{./}{figures/}}

\received{May 16, 2019}
\revised{June 7, 2019}
\accepted{June 11, 2019}
\submitjournal{ApJ}

\shorttitle{Ly$\alpha$ Halos Around $z\sim6$ Quasars}
\shortauthors{Drake et al., }

\begin{document}
\title{Ly$\alpha$ Halos Around $z\sim6$ Quasars}

\correspondingauthor{Alyssa Drake}
\email{drake@mpia.de}

\author[0000-0002-0786-7307]{Alyssa B. Drake}
\affil{Max Planck Institute for Astronomy, Konigstuhl, Heidelberg, Germany}
\author[0000-0002-0786-7307]{Emanuele Paolo Farina}
\affil{Max Planck Institute for Astronomy, Konigstuhl, Heidelberg, Germany}
\affil{Max Planck Institute for Astrophysics, Karl-Schwarzschild-Str, Garching, Germany}
\author[0000-0002-0786-7307]{Marcel Neeleman}
\affil{Max Planck Institute for Astronomy, Konigstuhl, Heidelberg, Germany}
\author[0000-0002-0786-7307]{Fabian Walter}
\affil{Max Planck Institute for Astronomy, Konigstuhl, Heidelberg, Germany}
\author[0000-0002-0786-7307]{Bram Venemans}
\affil{Max Planck Institute for Astronomy, Konigstuhl,  Heidelberg, Germany}
\author[0000-0002-0786-7307]{Eduardo Banados}
\affil{Max Planck Institute for Astronomy, Konigstuhl,  Heidelberg, Germany}
\author[0000-0002-0786-7307]{Chiara Mazzucchelli}
\affil{European Southern Observatory, Alonso de Cordova 3107, Vitacura, Region Metropolitana, Chile}
\author[0000-0002-0786-7307]{Roberto Decarli}
\affil{INAF - Osservatorio Astronomico di Bologna, Via Piero Gobetti, 93/3, 40129 Bologna BO, Italy}
 
\begin{abstract}
We present deep MUSE observations of five quasars within the first Gyr of the Universe ($z\gtrsim6$), four of which display extended Ly$\alpha$ halos. After PSF-subtraction, we reveal halos surrounding
two quasars for the first
time, as well as confirming the presence of two more halos for which tentative
detections exist in long-slit spectroscopic observations and narrow-band imaging. The four
Ly$\alpha$ halos presented here are diverse in morphology and size, they each
display spatial asymmetry, and none are centred on the position of the
quasar. Spectra of the diffuse halos demonstrate that none are dramatically
offset in velocity from the systemic redshift of the quasars ($\Delta$ v $< 200$
kms$^{-1}$), however each halo shows a broad
Ly$\alpha$ line, with a velocity width of order $\sim1000$ kms$^{-1}$. Total Ly$\alpha$ luminosities range between
$\sim$\,{\num{2E43}}\, erg s$^{-1}$ 
and $\sim$\,{\num{2E44}}\, erg s$^{-1}$, reaching 
maximum radial extents of $13 - 30$ pkpc from the quasar positions. We find larger sizes and higher Ly$\alpha$ luminosities than
previous literature results at this redshift, but find no correlation between the quasar properties and the Ly$\alpha$ halo, suggesting that the detected emission is most closely related to the physical properties of the circum-galactic medium. 
\end{abstract}


\section{Introduction} 
\label{Sect:intro}
 
The importance of studying the gas
immediately surrounding galaxies
has long been understood. In particular, as the circum-galactic medium (CGM) lies at the
interface between galaxies
themselves, and the diffuse hydrogen in the intergalactic medium (IGM) it holds the key to understanding some of the most fundamental concepts of galaxy
formation and evolution (e.g. \citealt{Tumlinson2017}). Until recently, the most prudent method of studying
this diffuse medium relied on absorption features imprinted on the spectra of
distant quasars by intervening gas e.g. \citealt{Hennawi2006a}, \citealt{HennawiProchaska2007}, \citealt{Farina2013} and \citealt{Farina2014} who each used this technique to study the CGM around other quasars along the line of sight.

Studying the CGM in emission however is a greater challenge. Nonetheless, in recent years the detection of Ly$\alpha$ halos around quasars at lower redshift (e.g. $3 \le z \le 4$) has become almost routine, in large part thanks to advances in instrumentation such as ESO's Multi-Unit Spectroscopic Explorer (MUSE; \citealt{Bacon2010}) on the Very Large Telescope (VLT). MUSE has allowed us to enter the paradigm whereby extended Ly$\alpha$ emission is ubiquitous not only around active galaxies (\citealt{Cantalupo2014}, \citealt{Borisova2016}, \citealt{ArrigoniBattaia2018}) but also normal star-forming galaxies out to $z\sim6$ (\citealt{Wisotzki2016}, \citealt{Leclercq2017}, \citealt{Drake2017a, Drake2017b}). 

The study of quasars in the first Gyr of the Universe ($z\gtrsim6$), is a probe through which we can directly observe young galaxies and their black holes during a rapid growth period. In addition, ionising radiation from the active galactic nucleus (AGN) can actually aid in the detection of CGM gas by causing it to shine more brightly in Ly$\alpha$. Furthermore, as large amounts of gas are directly funnelled onto the quasars' black holes, the accretion disks are luminous enough to allow their detection out to at least $z=7.5$ \citep{Banados2018} with the current facilities, allowing us to probe sources of ionising photons, and the gas fuelling their growth, well into the epoch of reionisation.

To date, a handful of detections of extended Ly$\alpha$ emission around very high redshift quasars ($z\sim6$) have been reported. \cite{Roche2014} presented long-slit spectroscopy of the radio-loud quasar \Roche\ at a redshift of $z=5.903$ (see also \citealt{Zeimann2011}). At present, this remains the highest-redshift radio-loud quasar to show signs of extended Ly$\alpha$ emission. The $z=6.43$ quasar \CFHQS\ \citep{Goto2009} is another example for which multiple
measurements of a Ly$\alpha$ halo have been reported in the literature. Following its
detection in narrow-band imaging, the halo was spectroscopically confirmed
in \cite{Willott2011} and \cite{Goto2012} and in these long-slit studies, the halo around \CFHQS\ appeared very similar to \Roche\ in terms of size and luminosity. Finally, \cite{Farina2017} targeted the $z=6.61$ QSO J$0305-3150$ with MUSE. With one of the highest SFRs and Eddington ratios above $z\sim6$, this quasar provides an ideal source to target a large gas reservoir surrounding one of the first QSOs. \cite{Farina2017} detected a faint Ly$\alpha$ halo around J$0305-3150$ extending $\sim 9$ pkpc, offset in velocity by $155$ km s$^{-1}$, and with a total luminosity of $(3.0 \pm 0.4) \times 10^{42}$ erg s$^{-1}$. 

\begin{table*}
	\centering
	\caption{QSOs for which archival MUSE data are presented in this paper. In the first four columns we list the object name, coordinates, and the reference for the discovery spectrum of each object. In the fifth column we list the current best estimate of each quasar's systemic redshift, in column 6 we give the method via which the systemic redshift was measured, and finally in column 7 we give the appropriate literature reference for the redshift measurement.}
	\label{tab:QSOs}
	\begin{tabular}{lcccccr} 
		\hline \hline
		{\bf{QSO}} & {\bf{RA}} & {\bf{DEC}} & {\bf{Discovery}} & {\bf{z}} & {\bf{z method}} & {\bf{z reference}} \\
		\hline
		{\bf{\Roche}} & 22:28:43.535  & +01:10:32.2 & \cite{Zeimann2011} & 5.903 & Ly$\alpha$ & \cite{Roche2014} \\
		{\bf{\Comp}} & 21:00:54.616 & -17:15:22.50 & \cite{willott2010a} & 6.081 & [CII] & \cite{Decarli2018} \\
		{\bf{\Italian}}  & 10:30:27.098 & +05:24:55.00 & \cite{Fan2001a}& 6.308 & \ion{Mg}{2} & \cite{Kurk2007}\\
		{\bf{\CFHQS}} & 23:29:08.275 & -03:01:58.80 & \cite{Willott2007} & 6.417 & \ion{Mg}{2} & \cite{Willott2011}\\
		{\bf{\Ffav}} & 15:26:37.841 & -20:50:00.66 & \cite{Mazzucchelli2017b} &6.587 & [CII] &  \cite{Decarli2018}\\
		\hline
	\end{tabular}
\end{table*}

Motivated by these studies, we search for Ly$\alpha$ halos in the deepest available MUSE observations of the highest redshift quasars ($z\sim6$). To reach approximately the surface brightness (SB) level achieved in \cite{Farina2017}, we queried the ESO archive for observations of exposure time $\sim 2$ hours or greater, resulting in a total sample of \noobjs\ objects listed in Table \ref{tab:QSOs}. SB levels in the wavelength layer where Ly$\alpha$ emission peaks range between \SBlimLyaItalian\ \SBunits\ and \SBlimLyaFfav\ \SBunits (see Table \ref{tab:tot_fluxes} for details). The observations include two objects with evidence for extended Ly$\alpha$ emission from long-slit spectra (\citealt{Willott2011}, \citealt{Goto2012}, \citealt{Roche2014}) or narrow-band imaging \citep{Momose2018}. 

Our analysis proceeds as follows; in Section \ref{Sect: Notes on individual objects} we provide context and notes on previous observations of quasars in our study, and in Section \ref{Sect: MUSE obs and DR} we describe our data reduction procedure and PSF-subtraction technique. We present our results in Section \ref{Sect: Results}, beginning with MUSE images and spectra of each QSO. We then show our PSF-subtracted data and estimate total fluxes of Ly$\alpha$ halos where present, and maximum extents. We next analyse the datacubes to a uniform surface brightness limit, and assess the morphology, spatial offsets, velocity offsets and integrated line profiles. Finally we include spatially resolved analyses of the kinematic structure of each halo. In Section \ref{Sect: Discussion} we discuss our measurements within the context of the literature, and speculate on the presence of evolution in the CGM, and the mechanisms powering the Ly$\alpha$ emission.  \\

Throughout this paper we use the terms `quasar' and `QSO' interchangeably, and assume a $\Lambda$CDM cosmology with $\Omega_m = 0.3$, $\Omega_{\Lambda} = 0.7$ and $h=70$ kms$^{-1}$.

\section{Previous Observations}
\label{Sect: Notes on individual objects}

In Table \ref{tab:QSOs} we summarise from the literature some basic properties of the QSOs analysed here.
\subsection{\Roche}
\Roche, at $z=5.903$, was discovered via its radio emission as part of the Stripe82 VLA survey (peak flux density 0.31 mJy; \citealt{Zeimann2011}). \Roche\ is optically faint however, with {\mbox{M$_{{\rm{1450}}}$ $= -24.53$}}, and consequently its radio loudness parameter is high, R $\sim 1100$. With an LRIS spectrum, \cite{Zeimann2011} measured the velocity width of the Ly$\alpha$ line in the QSO spectrum to be $1890$ km s$^{-1}$. \cite{Roche2014} returned to \Roche\ with GTC-OSIRIS, and were the first to detect signs of an extended Ly$\alpha$ halo around the quasar. They measured a velocity width of the Ly$\alpha$ line (including QSO) of $1189 \pm 24$ km s$^{-1}$. In addition they extracted an offset spectrum towards the South-East of the quasar, and measured a flux of $2.02 \pm 0.46 \times 10^{-17}$ \fluxunits, with a peak of emission at $8386.26 \pm 2.41$ \AA. The Ly$\alpha$ luminosity of the halo was constrained to be L(Ly$\alpha$) $\leq$ \num{7.8E42} \lumunits\ extended for ``at least" $9$ pkpc. \Roche\ is the only known radio-loud object in this sample. MUSE data for this object total 11.30 h (PI: Roche).

\subsection{\Comp}
\Comp\ was uncovered as part of the Canada-France-High-$z$-Survey and reported in \cite{Willott2010}, it has an {\mbox{M$_{{\rm{1450}}}$ $= -25.55$}} \citep{Banados2016}. It was also targeted as one of $27$ $z \sim 6$ quasars in the [CII] survey of \cite{Decarli2018}, allowing for an accurate determination of the systemic redshift. MUSE data were taken for this object (PI: Decarli), amounting to 3.70 h in total.

\subsection{\Italian}
\Italian\ at $z = 6.304$ \citep{Fan2001}
was one of the earliest quasars to be discovered at $z \sim 6$ in the Sloan Digital Sky Survey (SDSS), and has an {\mbox{M$_{{\rm{1450}}}$ $= -26.99$}} \citep{Banados2016}. \cite{Momose2018} found no halo for this quasar in their Subaru narrow-band imaging. MUSE data are of integration time 6.43 h (PI: Karman).

\subsection{\CFHQS}
\CFHQS\ ($z=6.417$) was discovered as part of the Canada-France High-$z$ Quasar Survey (CFHQS; \citealt{Willott2007}), and has {\mbox{M$_{{\rm{1450}}}$ $= -25.25$}} \citep{Banados2016}. \cite{Goto2009} were first to find evidence of an extension to the Ly$\alpha$ emission, tentatively seen in Subaru/Suprime-Cam narrow-band imaging - they estimated an extent of $4$ arcsec on the sky. Presence of the halo was confirmed in the long-slit spectroscopy of \citealt{Willott2011} (measuring $8 \times 10^{43}$ \lumunits\ over $15$ pkpc or $2.7$ arcsec) and \citealt{Goto2012} ($1.7 \times 10^{43}$ \lumunits\ over $4$ arcsec), and subsequently further narrow-band imaging in \citealt{Momose2018} ($1.5 \times 10^{44}$ \lumunits\ over $6.9$ arcsec in diameter). This quasar was observed as part of MUSE Science Verification, with an integration time of 2 h.

\subsection{\Ffav}
\Ffav\ is the highest redshift object in our study, at $z=6.586$, discovered in \cite{Mazzucchelli2017b}. This quasar is one of the brightest known at $z > 6.5$; M$_{{\rm{1450}}}$ $= -27.20$ \citep{Banados2016}; and is known to have a very close companion galaxy detected in [CII], and reported in \cite{Decarli2018}. MUSE data were taken as part of a programme (PI: Farina), with a total integration time of 3.20 h. 

\section{MUSE Observations and analysis}
\label{Sect: MUSE obs and DR}

We analysed the deepest available MUSE data for quasars at $z\sim6$. A summary of the sample is given in Table \ref{tab:QSOs}. 

\subsection{Data Reduction}
MUSE data and raw calibrations were downloaded from the ESO archive for each of the quasars listed in Table \ref{tab:QSOs}. We used the MUSE Data Reduction Software (version 2.4.1) to reduce each individual exposure, combining them into deep datacubes for each object. Once the data were fully reduced we ran the Zurich Atmospheric Purge software (ZAP; \citealt{Soto2016}) to perform additional sky subtraction. On these final `ZAPped' datacubes we perform the remainder of our analysis\footnote{(We note that in parallel to this analysis we assessed the ``un-ZAPped" data cubes, finding consistent results.)}. \\

\begin{figure*}
  \centering
	\includegraphics[width=\textwidth]{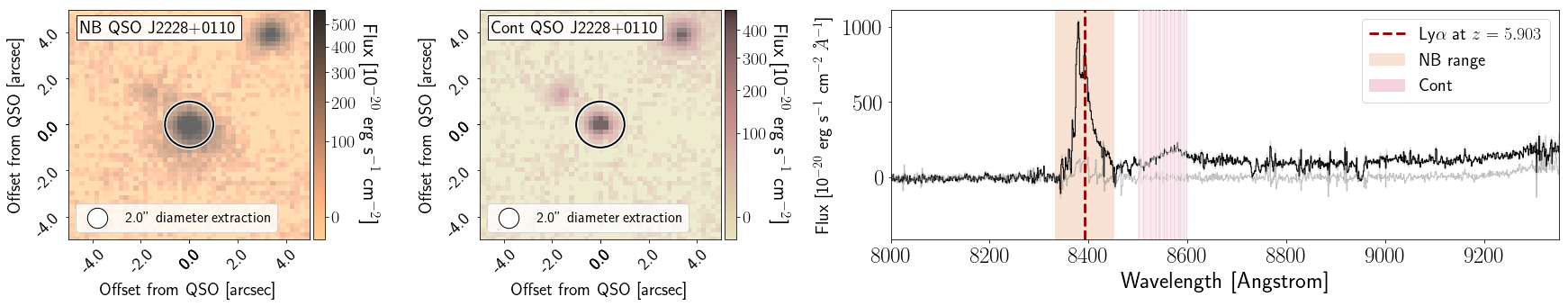}
	\includegraphics[width=\textwidth]{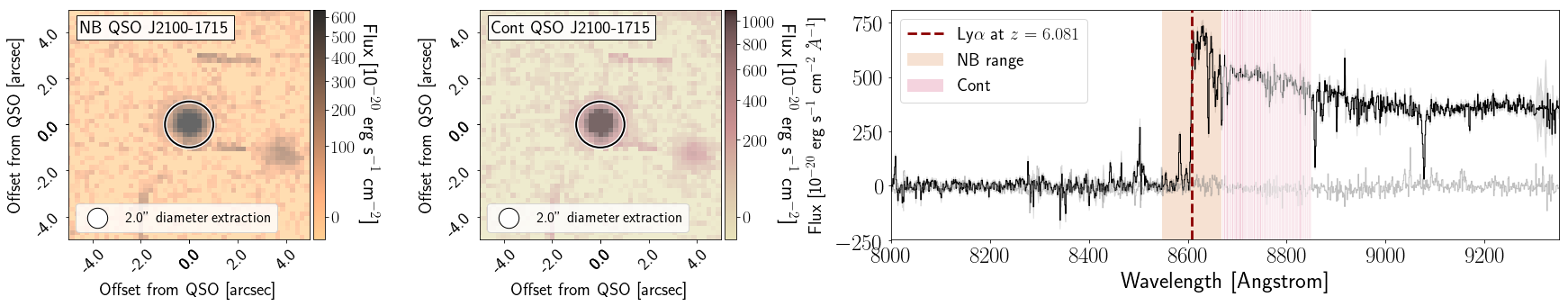}
	\includegraphics[width=\textwidth]{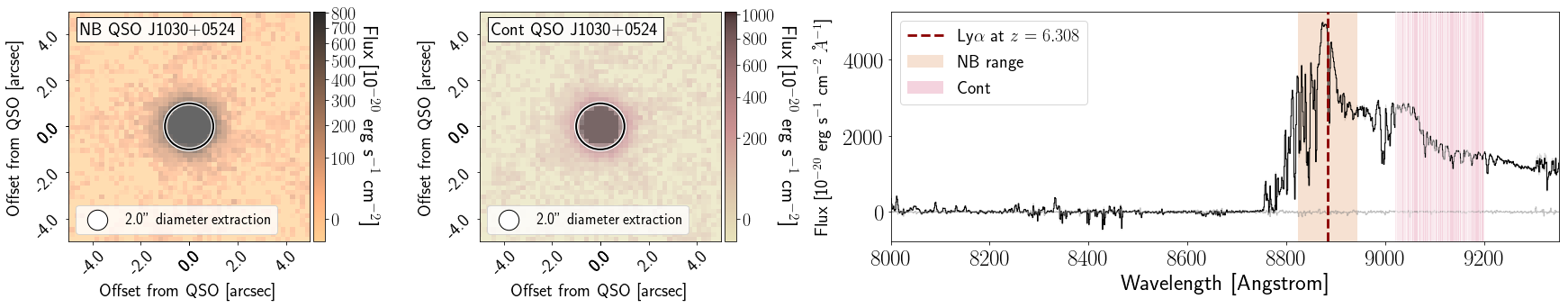}
	\includegraphics[width=\textwidth]{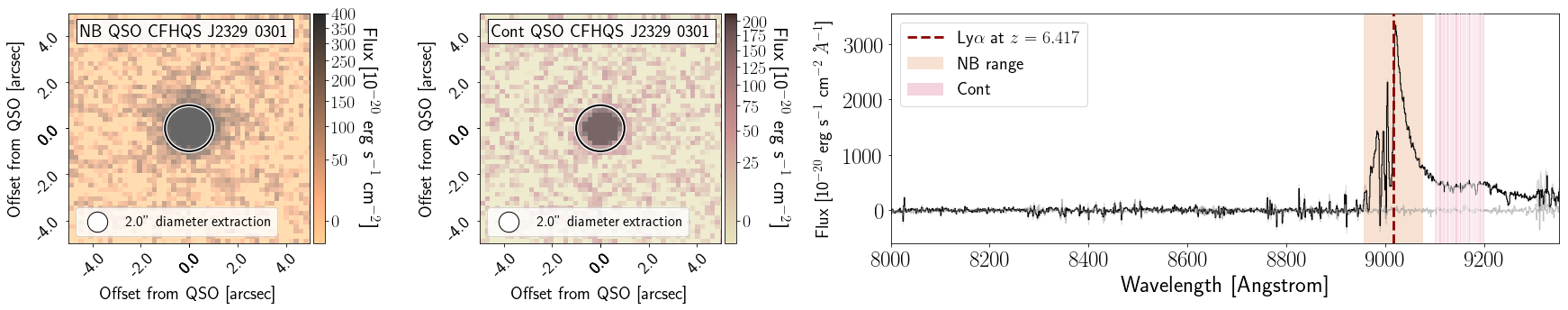}
	\includegraphics[width=\textwidth]{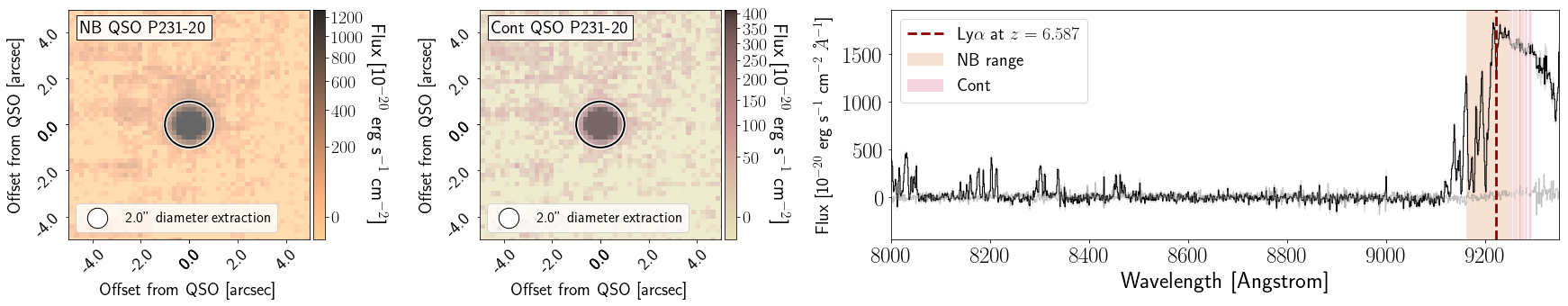}
    \caption{MUSE images and spectra of the \noobjs\ $z\sim6$ QSOs analysed here. The left-hand panels show fixed-width narrow-band images of $\Delta \lambda = 120$ \AA\ centred on the predicted wavelength of Ly$\alpha$ from our best-estimate of the systemic redshift (see Table \ref{tab:QSOs}). The wavelength range of each narrow-band image is highlighted by the orange region on the spectrum. In the central panels we show the `clean' continuum images, i.e. collapses of the QSO continuum emission, constructed using wavelength layers of the cube where night sky emission is low (see Section \ref{Sect:PSF} for details). The layers combined to make the image are highlighted in pink on the spectrum. In the final column, we show the QSO spectrum extracted in an aperture of 2 arcseconds in diameter, the extraction region is given by the black circle on the narrow-band and continuum-collapse images. Dashed red lines on the spectrum give the predicted peak wavelength of the Ly$\alpha$ emission according to the systemic redshift.} 
	\label{Fig:1_im_spec}
\end{figure*}

\subsection{PSF Subtraction}
\label{Sect:PSF}

The detection of low-surface brightness emission around a bright point source requires removing the contribution of light from the unresolved object. This is achieved by characterising the point-spread function (PSF) in the data, typically using a bright star in the field, or in the case of a quasar, using the spectral continuum from the quasar itself. The quasar continuum is dominated by light from the accretion disk of the AGN, which is an unresolved source at the resolution of MUSE. 
For each QSO in this work we construct our model of the PSF following the steps outlined below. We utilise the simultaneous spatial and spectral information in the MUSE datacube to collapse several spectral layers of the quasar continuum, creating a ``PSF image". Using the quasar itself allows us to avoid issues related to PSF variations across the field, and spatial interpolation. Indeed, in the analysis of the $z\sim6$ quasar presented in \cite{Farina2017}, these two commonly-applied PSF subtraction techniques were tested, and found to be equally reliable. 

In constructing our PSF-image, there are two considerations in producing the most reliable result. Firstly, the need for signal to noise (S/N) in the PSF image means including as wide a spectral range as possible, however at these wavelengths (particularly $>7000$ \AA) emission from the night sky varies rapidly as a function of wavelength, resulting in differences in the noise properties of adjacent wavelength layers even after careful sky-subtraction. This means that the S/N in the PSF image is actually improved by excluding wavelength layers of the cube which are most affected by sky lines. 

The wavelength layers chosen to construct the PSF image for each quasar are highlighted in pink on the spectrum shown in Figure \ref{Fig:1_im_spec}\footnote{We note that in the case of \Ffav, the PSF image was by necessity constructed using wavelength layers of the cube which could contain the high-velocity tails of Ly$\alpha$ emission. This might lead to a slight over-subtraction of the halo and therefore our flux estimate is a conservative one.}. We then work systematically through each wavelength layer of the cube, scaling the flux in the peak pixel of the PSF image to the flux of the QSO in the same spatial pixel, and subtracting the scaled-PSF-image from the datacube wavelength layer. This way we produce an entire PSF-subtracted datacube\footnote{To test the reliability of the PSF-image scaling we also conducted a test using the average value of 5 pixels in the centre of the PSF image scaled to that of 5 pixels at the centre of the QSO, and verified that our results did not change.}. 

Once the PSF-subtracted datacube has been created, we mask an ellipse of dimensions equal to the Gaussian-equivalent PSF (the FWHM of a 2-dimensional Gaussian fit to the PSF image) in each wavelength layer, and exclude this central region from further analysis since it is likely to contain complex residuals.

\section{Results}
\label{Sect: Results}
 We present the MUSE data for each QSO, together with the analysis of the PSF-subtracted datacubes in the subsections below.
 
\subsection{Images and Spectra}
In Figure \ref{Fig:1_im_spec} we show MUSE images and spectra for each QSO. The first panel shows a narrow-band image ($\Delta\lambda=120$\AA) centred on the predicted position of the Ly$\alpha$ line from our best estimate of the quasar systemic redshift. In the second panel we show the `PSF image' constructed as described in Section \ref{Sect:PSF} for the QSO continuum, and in the third panel we show the quasar spectrum extracted in an aperture of $2$ arcseconds in diameter. The wavelength range of each image is highlighted on the spectrum, and likewise the extraction aperture of the spectrum is shown on each image.\\

\subsection{Ly$\alpha$ halos at $z\sim 6$}
\label{Sect:Halos}
We detect extended Ly$\alpha$ emission around \nohalos\ out of \noobjs\ quasars in the PSF-subtracted datacubes. We proceed now by estimating the total fluxes of the halos, and maximum extents that the diffuse emission reaches from the positions of the quasars. In order to correctly measure the extent and luminosity of the Ly$\alpha$ halos, we need to cover the full spectral width of the Ly$\alpha$ line. As a result, for each quasar our pseudo narrow-band is comprised of a different spectral range, and hence reaches a different surface brightness limit. Wavelength layers which comprise each of these narrow bands are shown in Appendix A, and their corresponding 1$\sigma$ surface brightness limits are listed in Table \ref{tab:tot_fluxes}. In order to facilitate comparison between the halos we will also analyse the data to a uniform surface brightness limit in Section \ref{Sect:UniSB} to assess the morphology, spatial offset of the emission peak from the quasar and the integrated velocity offset and velocity widths of the halos. Finally, in Section \ref{Sect:ResKinematics} we employ a S/N cut {\emph{per voxel}} in order to produce the most sensitive maps of the kinematic structure of each halo, details follow in Section \ref{Sect:ResKinematics}.

\begin{figure*}
  \centering
	\includegraphics[width=0.6\textwidth]{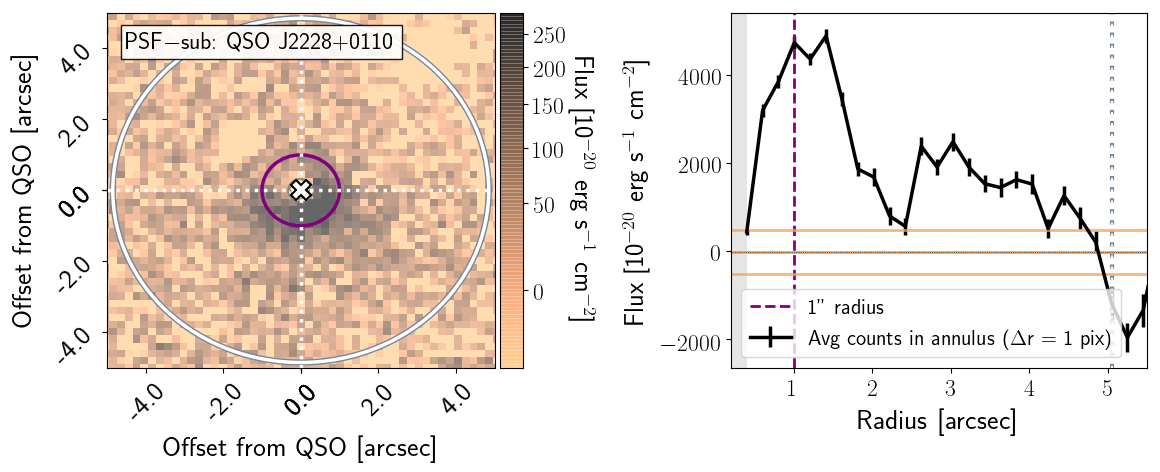}
	\includegraphics[width=0.6\textwidth]{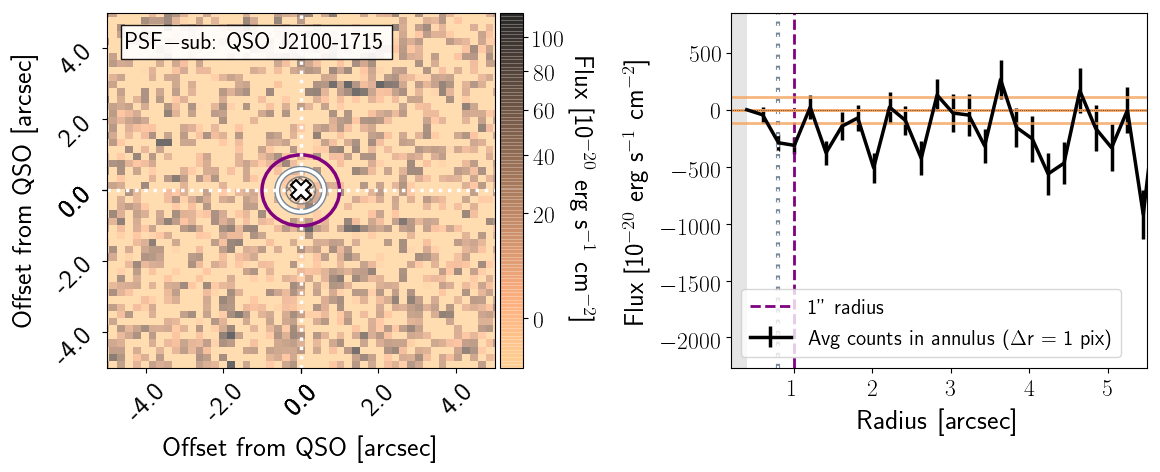}
	\includegraphics[width=0.6\textwidth]{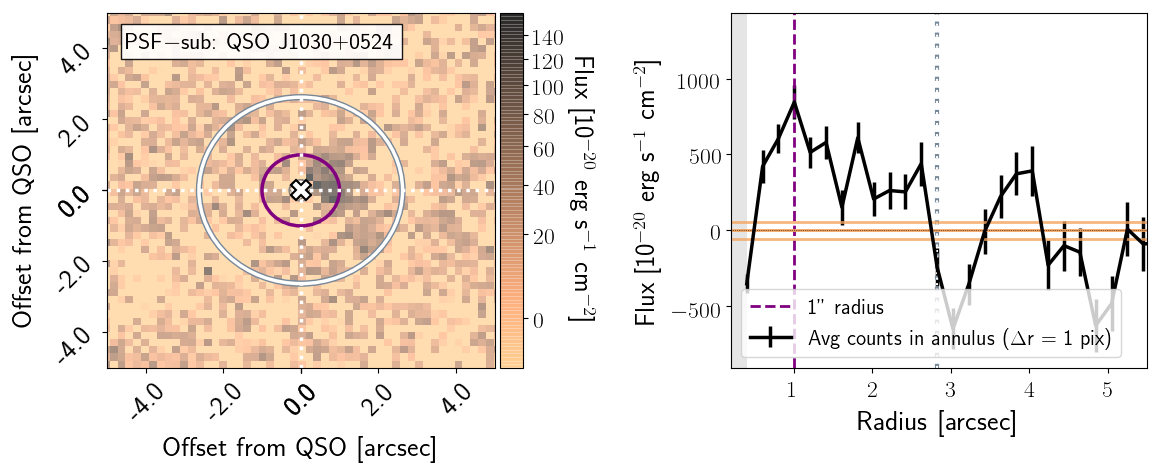}
	\includegraphics[width=0.6\textwidth]{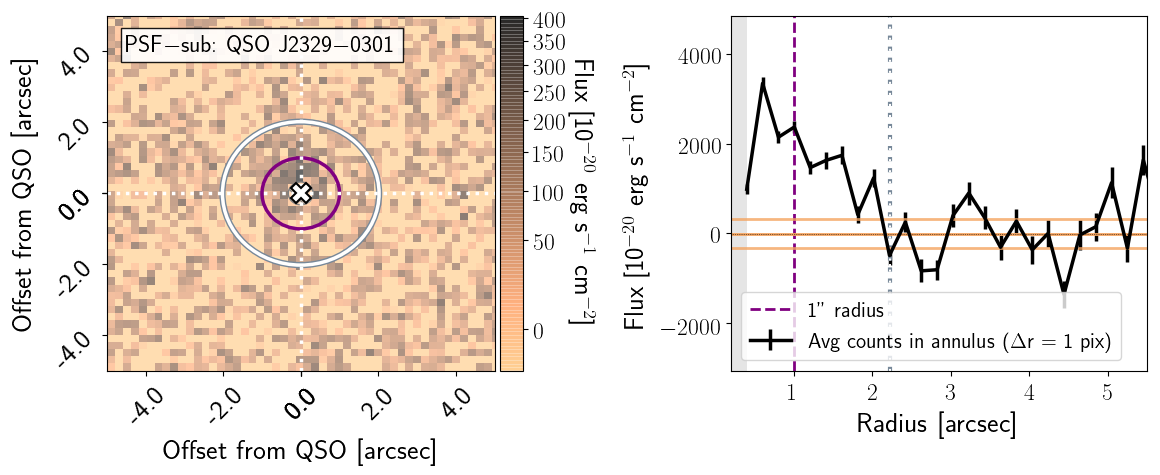}
	\includegraphics[width=0.6\textwidth]{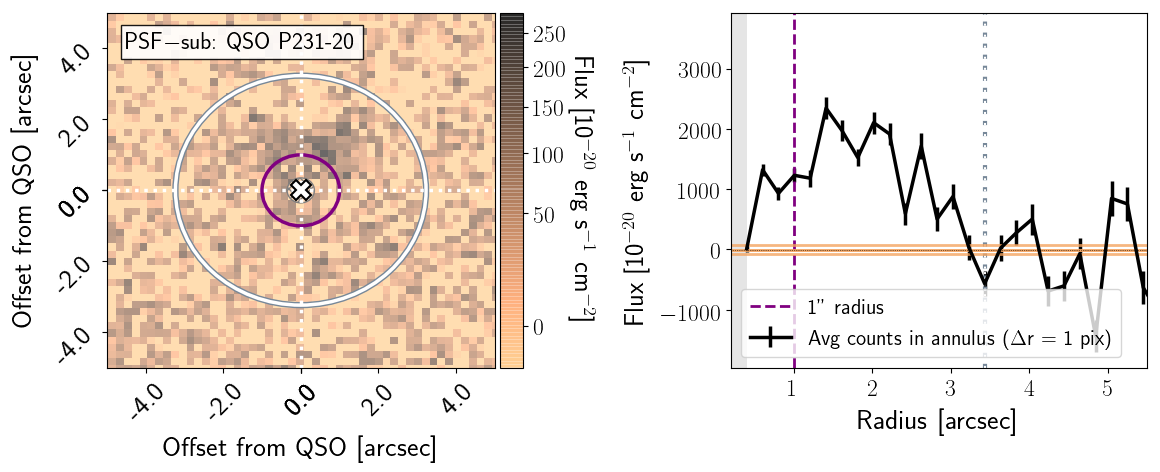}
    \caption{Fixed-width narrow-band images centred on the Ly$\alpha$ line, and azimuthally-averaged radial light profiles. Images in the left-hand column are constructed such that they encompass the spectral width of the Ly$\alpha$ line in the PSF-subtracted cube, details are outlined in Section \ref{Sect:Fluxes} and shown in Appendix A. Overlaid is an aperture of 2 arcseconds in diameter (purple circle) and an aperture to show the profile radius determined in the adjacent column of panels. On the right-hand side we show each light profile and its 3 $\sigma$ error out to a radius of $\sim 5$ arcseconds, and exclude the very central region of the profiles which cannot be trusted due to possible PSF subtraction artefacts (shaded grey). The purple dashed line corresponds to the 2-arcsecond diameter aperture, and the grey dashed line indicates the radius where the light profile hits the background of the observations (dark orange lines represent the background level, and light orange lines denote the 1$\sigma$ deviation).} 
	\label{Fig:tot_flux}
\end{figure*}

\subsubsection{Total Ly$\alpha$ fluxes and maximum extents}
\label{Sect:Fluxes}

We begin by visually inspecting the PSF-subtracted datacubes, and extracting a spectrum in a large aperture chosen by eye to encapsulate the visible extended emission (typically 6 arcseconds in diameter, see Appendix A). From this spectrum we then choose the appropriate spectral window over which to collapse the cube to encapsulate the entire spectral width of the Ly$\alpha$ line, resulting in our preferred pseudo narrow-band image.

In accordance with other works in the literature (e.g. \citealt{Borisova2016}, \citealt{ArrigoniBattaia2018}, \citealt{Wisotzki2016}, \citealt{Drake2017a}, \citealt{Drake2017b}) we use azimuthally-averaged profiles on these fixed-width narrow-band images in order to track the curve-of-growth of the extended emission around each object, and sum the Ly$\alpha$ flux within the radius where the light profile hits the background of the observations. Our best estimates of the total flux of each PSF-subtracted halo are presented in Table \ref{tab:tot_fluxes}, together with the wavelength layers over which the estimate was made, the maximum extent of the emission, and the corresponding total Ly$\alpha$ luminosity of each halo. We display the narrow-band images, and light profiles in Figure \ref{Fig:tot_flux}.

 By far the most prominent halo is that around radio-loud \Roche, extending \RocheRadArcsec\ or \RocheRadPkpc, making it to date the largest halo detected around a $z \sim 6$ quasar. We measure the total flux of the halo as \RocheTotFlux\ \fluxunits\ across this extent, a larger flux than any of the existing literature measurements. \citealt{Roche2014} for example measured a halo flux of $2.02 \pm 0.46 \times 10^{-17}$ \fluxunits, thus our estimate is an order of magnitude larger than previous measurements. This is likely to be due to very different observational techniques  - \citealt{Roche2014} estimate the halo flux in a long slit offset from the position of the quasar, confirming the presence of extended emission, but by no means an observation designed to precisely measure the halo flux.

 \Comp, the next highest redshift object shows no sign of extended emission above the SB limit in these data (\SBlimNBComp\ \SBunits) in our narrow-band collapse around the Ly$\alpha$ line. 
 
 \Italian\ shows a small but distinct halo towards the West extending \ItalianRadArcsec\ or \ItalianRadPkpc\ from the position of the quasar. Within this radius we measure a total flux of \ItalianTotFlux \fluxunits.

 \CFHQS\ is the only quasar other than \Roche\ to have previous measurements of an extended Ly$\alpha$ halo. We detect a halo much as \cite{Goto2009}, \cite{Willott2011}, \cite{Goto2012} and \cite{Momose2018} do, extending North. In our data the halo reaches a maximum radial extent of \CFHQSRadArcsec\ or \CFHQSRadPkpc, with a total halo flux of \CFHQSTotFlux \fluxunits. Interestingly, our halo luminosity estimate is very similar to that of \cite{Momose2018}, although \cite{Momose2018} find a larger extension on sky (6.9 arcseconds in diameter in an image reaching an SB limit of {\num{4E-18}} \SBunits at the $3 \sigma$ level.). These differences are probably due to the image produced in this work being sensitive only to a higher SB limit (e.g. a $1 \sigma$ limit of \SBlimNBCFHQS \SBunits as quoted in Table \ref{tab:tot_fluxes}) meaning the halo is less-well detected in its outskirts, but conversely the ability of MUSE to encapsulate the spectral width of the line very precisely (compared to narrow-band flux losses) means that we recover approximately the same total flux/luminosity.

 Finally, we see a Ly$\alpha$ halo around \Ffav\ extending \FfavRadArcsec, or \FfavRadPkpc, of similar extent to \Italian, but with a total flux of \FfavTotFlux\ \fluxunits, approximately twice the flux of \Italian\ (making it comparable in flux to \CFHQS). 
 In Table \ref{tab:tot_fluxes} we report our best estimate of the maximum radial extent that each halo reaches from the position of the quasar, hitting SB background levels between \SBlimNBItalian\ \SBunits\ and \SBlimNBFfav\ \SBunits. All \nohalos\ of the halos have total Ly$\alpha$ luminosities greater than L$_{{\rm{Ly\alpha}}} =$ \num{1E43} \lumunits, we will return to the halo luminosities in the context of other works in Section \ref{Sect: Discussion}.

\begin{table*}
	\centering
	\caption{Total fluxes and luminosities for the \noobjs\ QSOs discussed in this work. In the first column we give the QSO names. In the second and third columns we report our measurement of the total flux and corresponding log luminosity. These are the values summed within our measurement of the maximum extent of the halos on-sky. Columns 4 and 5 report the maximum radial extent we measure for each QSO, in arcseconds, and pkpc at the respective QSO redshifts. In the sixth column we give the wavelength range and width (N layers) over which the measurements were made. In the final two columns we give the 1 $\sigma$ surface brightness limits first in the wavelength layer where the peeak of the Ly$\alpha$ emission is detected, and finally in the narrow-band image constructed from the wavelength range listed in column 6.}
	\label{tab:tot_fluxes}
	\begin{tabular}{lcccccccc} 
		\hline \hline
\multirow{3}{*}{{\bf{QSO}}} & \multirow{2}{*}{{\bf{Total Flux}}} & \multirow{2}{*}{{\bf{log LLya}}} & \multicolumn{2}{c}{\multirow{2}{*}{\bf{Maximum Extent}}} & {\bf{NB width}} & {\bf{SB Ly$\alpha$}} & {\bf{SB NB}}\\
 & {\multirow{2}{*}{[erg s$^{-1}$ cm$^{-2}$]}} & {\multirow{2}{*}{[erg s$^{-1}$]}}  & {\multirow{2}{*}{[arcsec]}} & {\multirow{2}{*}{[pkpc]}} & {\bf{(N)}} & {\bf{($1\sigma$)}} & {\bf{($1\sigma$)}}\\
   &   & 	 &  &  & [Angstroms] & \multicolumn{2}{c}{[\SBunits]}  \\ 
 
\hline
{\bf{\Roche}} & \RocheTotFlux & \RocheTotLum & \RocheRadArcsec & \RocheRadPkpc & 8355-8435 (\layersNBRoche) & \SBlimLyaRoche & \SBlimNBRoche  \\
{\bf{\Comp}} & $--$ & $--$ & $--$ & $--$  & 8600-8650 (\layersNBComp) & \SBlimLyaComp & \SBlimNBComp \\
{\bf{\Italian}}  & \ItalianTotFlux & \ItalianTotLum & \ItalianRadArcsec &  \ItalianRadPkpc & 8865-8900 (\layersNBItalian)& \SBlimLyaItalian & \SBlimNBItalian  \\
{\bf{\CFHQS}}  & \CFHQSTotFlux & \CFHQSTotLum & \CFHQSRadArcsec & \CFHQSRadPkpc & 9000-9045 (\layersNBCFHQS)& \SBlimLyaCFHQS & \SBlimNBCFHQS \\
{\bf{\Ffav}} & \FfavTotFlux & \FfavTotLum & \FfavRadArcsec & \FfavRadPkpc & 9210-9260 (\layersNBFfav) & \SBlimLyaFfav & \SBlimNBFfav \\
		\hline
	\end{tabular}
\end{table*}

\subsection{Properties of diffuse Ly$\alpha$ above a uniform surface-brightness limit}
\label{Sect:UniSB}
\begin{figure*}
  \centering
	\includegraphics[width=0.7\textwidth]{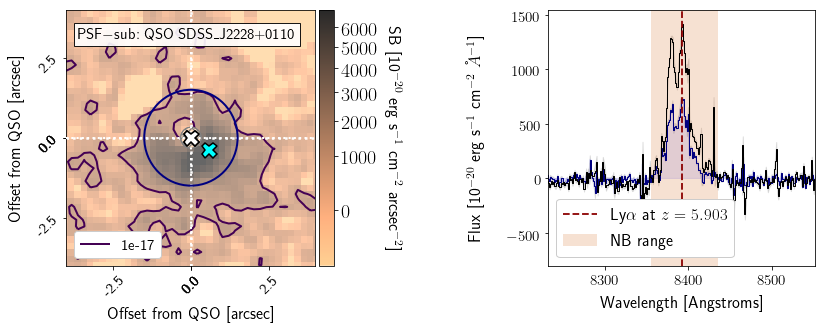}
	\includegraphics[width=0.7\textwidth]{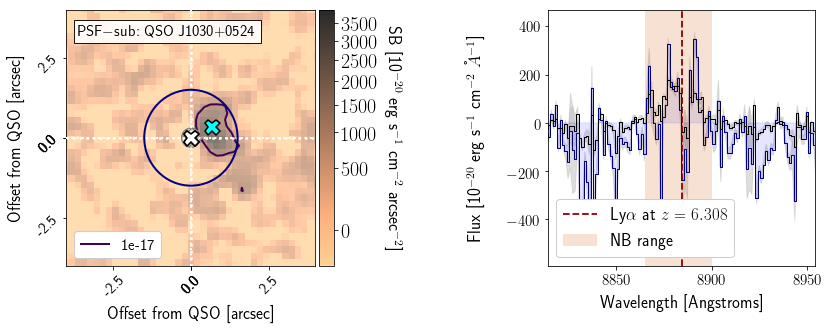}
	\includegraphics[width=0.7\textwidth]{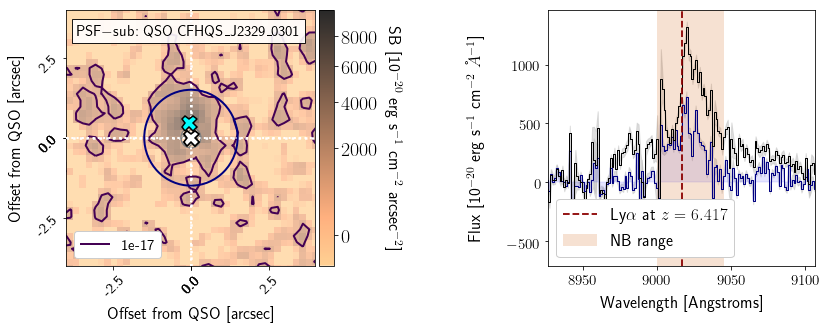}
	\includegraphics[width=0.7\textwidth]{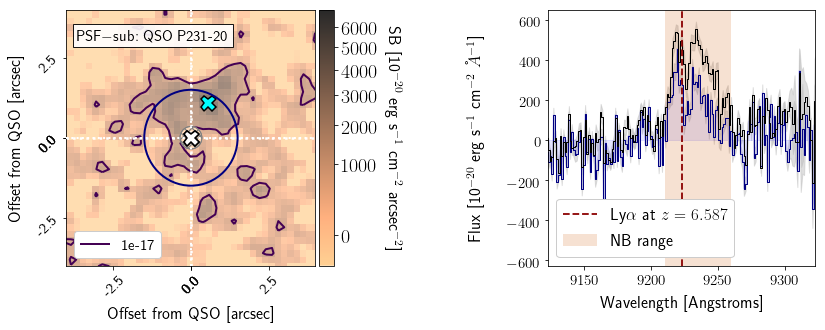}
    \caption{Smoothed surface brightness images and spectra of the PSF-subtracted Ly$\alpha$ halo. Note that the central 4 pixels (approximately the size of the PSF) have been masked in the datacube and thus do not contribute to the images or spectra shown here. In the left-hand panels we show the same narrow-band image as in Figure \ref{Fig:tot_flux}, now smoothed with a Gaussian kernel of $\sigma =$ \GaussianSmoothingKernelSize, in units of surface brightness, and contoured at a level of SB$=$ \contourhigh\ \SBunits\ (black line). On this image we show the quasar position (white cross) and the peak of the Ly$\alpha$ emission in the halo which remains after PSF-subtraction (cyan cross), in addition to an aperture of 2 arcseconds in diameter. In the right-hand panels we show in filled-blue the spectrum extracted from the PSF-subtracted cube within the 2-arcsecond diameter aperture, and overlay in black a spectrum extracted by summing all the voxels (volumetric pixels) lying within the \contourhigh\ \SBunits\ contour. The dashed red line gives the predicted position of the peak of the Ly$\alpha$ emission according to the systemic redshift of the quasar (Table \ref{tab:QSOs}), and the shaded orange region corresponds to the wavelength region that makes up the narrow-band image.} 
	\label{Fig:2_blobs}
\end{figure*}

In literature searches, authors frequently survey objects to a uniform depth and consider only emission above some signal to noise or surface brightness level. In the data analysed here, the observations reach various depths (summarised in Table \ref{tab:tot_fluxes}), and so in order to directly compare emission around the quasars we must compare the properties of the diffuse emission to a depth easily reached in all the observations. In Figure \ref{Fig:2_blobs} we present the same narrow-band images described in Section \ref{Sect:Fluxes}, now in surface brightness units, and smoothed with a Gaussian kernel of \GaussianSmoothingKernelSize\ to reveal the morphology of the halos. The images are contoured at \contourhigh\ \SBunits\ to highlight the brightest emission around each source.  
 
We discuss below our analysis of the morphology and emission-line properties of each halo, and summarise our findings in Table \ref{tab:diffuse17}.

\subsubsection{Nebula morphology and spatial offsets}
\label{Sect:MorphSpatOff}

It is evident from Figure \ref{Fig:2_blobs} that the extended emission around the \nohalos\ QSOs showing a Ly$\alpha$ halo is diverse in appearance and size. With the possible exception of \CFHQS, the halos are not centred on the spatial position of the QSOs, and none appear regular in shape. 

We compute for each object the spatial offset between the position of the QSO and the peak of the halo emission (see Table \ref{tab:diffuse17}). In Figure \ref{Fig:2_blobs} we overlay a white cross at the position of the QSO, and a cyan cross at the peak of the halo emission in each panel. The halos are on average offset by \avgOffsetArcsecs\ (\avgOffsetpkpc\ across their respective redshifts). Interestingly, the halo most removed from its associated QSO is radio quiet, and is the highest redshift object in our sample, \Ffav, which is also known to have a close companion \citep{Decarli2018}. 

We also list in Table \ref{tab:diffuse17} the area in square arcseconds of the \mbox{\gtSBbright} emission in this image.

\subsubsection{Velocity offsets and integrated line profiles}
\label{Sect:VelOff}

We consider the kinematic properties of the halos in two ways, beginning in this Section with the integrated properties of the halos. Firstly, in Figure \ref{Fig:3_line_profiles} we show spectra of the quasar (upper panels), and those of the diffuse halo emission after PSF-subtraction (lower panels). 

Spectra of the halos are extracted in apertures of $3$ arcseconds in diameter, but excluding the central region (PSF). 

Overall, the halos show little velocity offset from their respective QSOs, either with respect to the systemic redshift (or our best-estimate thereof c.f. Table \ref{tab:QSOs}) or the peak of the quasar's Ly$\alpha$ emission. The absence of a significant velocity offset between the halo and the QSO, suggests that we are not witnessing an infall/outflow scenario e.g. the predictions of {\cite{Villar-Martin2007} and {\cite{Weidinger2005}}.

Interestingly, the velocity widths of the emission line profiles for each halo are very broad, of only marginally lower velocity width than the Ly$\alpha$ from the QSO. This is an interesting contrast to the halos seen in \cite{Borisova2016}, however is strikingly similar to the case seen in \cite{Ginolfi2018}.

\begin{table*}
	\centering
	\caption{Properties of diffuse emission of SB $>$ \num{1E-17} \SBunits. In the first column we give the QSO names. In the second column we present the contiguous area in square arcseconds that is above SB $>$ \num{1E-17} \SBunits. In the third and forth columns we give the spatial offset of the peak of the halo emission from the position of the QSO, in arcseconds and pkpc. In the fifth column we give the velocity offset of the halo from the QSO, and in the final column we give the velocity width (FWHM) of the Ly$\alpha$ line arising from the halo.}
	\label{tab:diffuse17}
	\begin{tabular}{lcccccc} 
		\hline \hline
{\bf{QSO}}	& {\bf{Area $>$ \num{1E-17}}} & \multicolumn{2}{c}{\bf{Neb offset}}  &	  {\bf{Vel offset}} & {\bf{Vel}}\\
	&  [arcsec$^2$] &  [arcsec] &  [pkpc] &	[kms$^{-1}$] & [kms$^{-1}$]\\
\hline
{\bf{\Roche}}	& 18.16	& 0.69 & 4.0 & -126.25 & 1052.96 \\
{\bf{\Comp}} &	$--$ &	$--$ &	$--$  & $--$  & $--$ \\
{\bf{\Italian}} & 1.47 &  0.74 &  4.1 &  18.47 & 531.84 \\
{\bf{\CFHQS}} & 5.92 & 0.44 & 2.5 & 116.18 & 955.21 \\
{\bf{\Ffav}}	& 6.57 & 1.21 & 6.5 & 181.66 & 942.94 \\ 
		\hline
	\end{tabular}
\end{table*}

\begin{figure*}
\centering
	\includegraphics[width=0.48\textwidth]{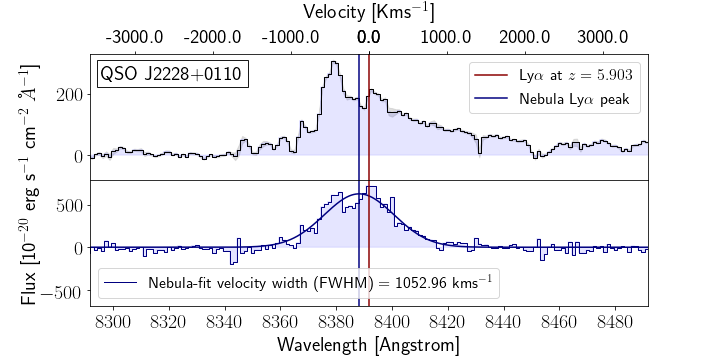}
	\includegraphics[width=0.48\textwidth]{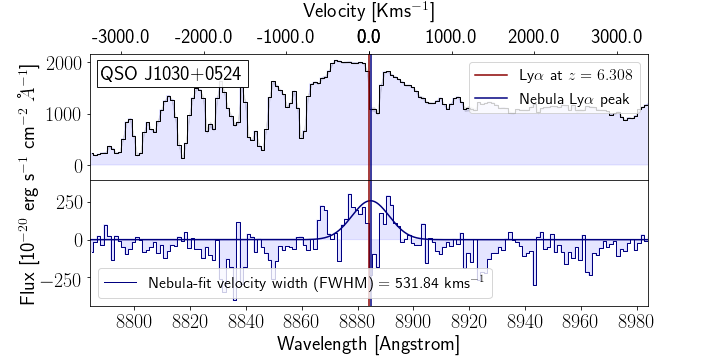}
	\includegraphics[width=0.48\textwidth]{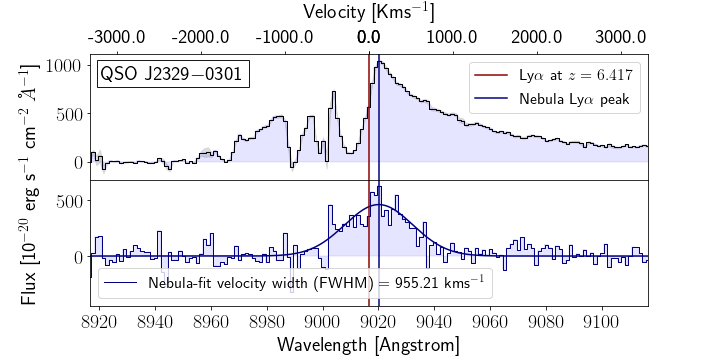}
	\includegraphics[width=0.48\textwidth]{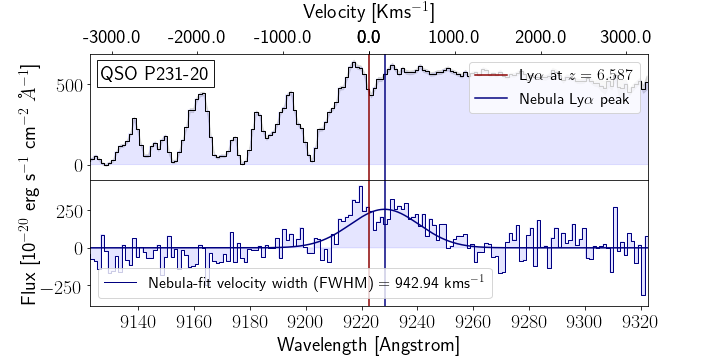}
    \caption{Spectra of each QSO, and its Ly$\alpha$ halo. In the upper panels we show the QSO spectrum extracted from the central four pixels (i.e. approximately the size of the PSF). In the lower panels we show spectra of the Ly$\alpha$ halo in the PSF-subtracted datacube in an aperture of 3 arcseconds in diameter. Note that for the Ly$\alpha$ halo spectra the central four pixels centred on the QSO have been masked from the datacube and do not contribute to the Ly$\alpha$ halo spectrum. Overlaid on each pair of spectra are the predicted position of the Ly$\alpha$ line according to our best estimate of the systemic redshift of the quasar (red line); and the peak of the PSF-subtracted emission (blue line). For each object we fit a Gaussian profile to the PSF-subtracted line to estimate the velocity width. The best-fit FWHM values in kms$^{-1}$ are noted on each panel.} 
	\label{Fig:3_line_profiles}
\end{figure*}

\begin{figure*}
  \centering
	\includegraphics[width=0.9\textwidth]{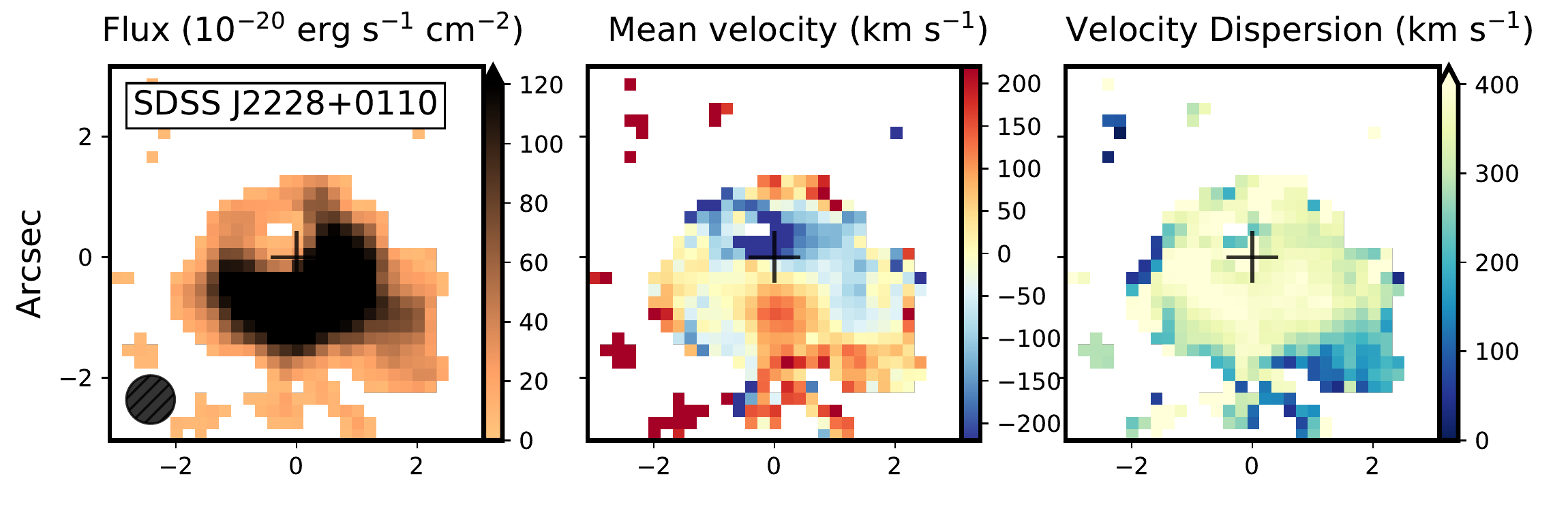}
	\includegraphics[width=0.9\textwidth]{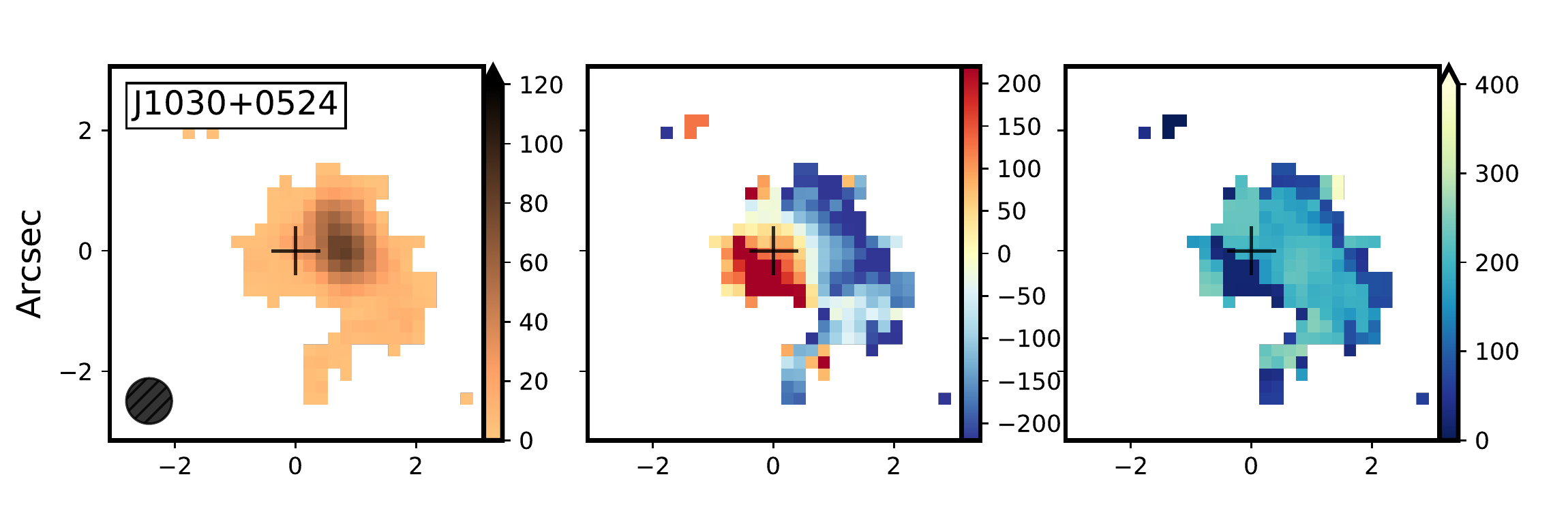}
	\includegraphics[width=0.9\textwidth]{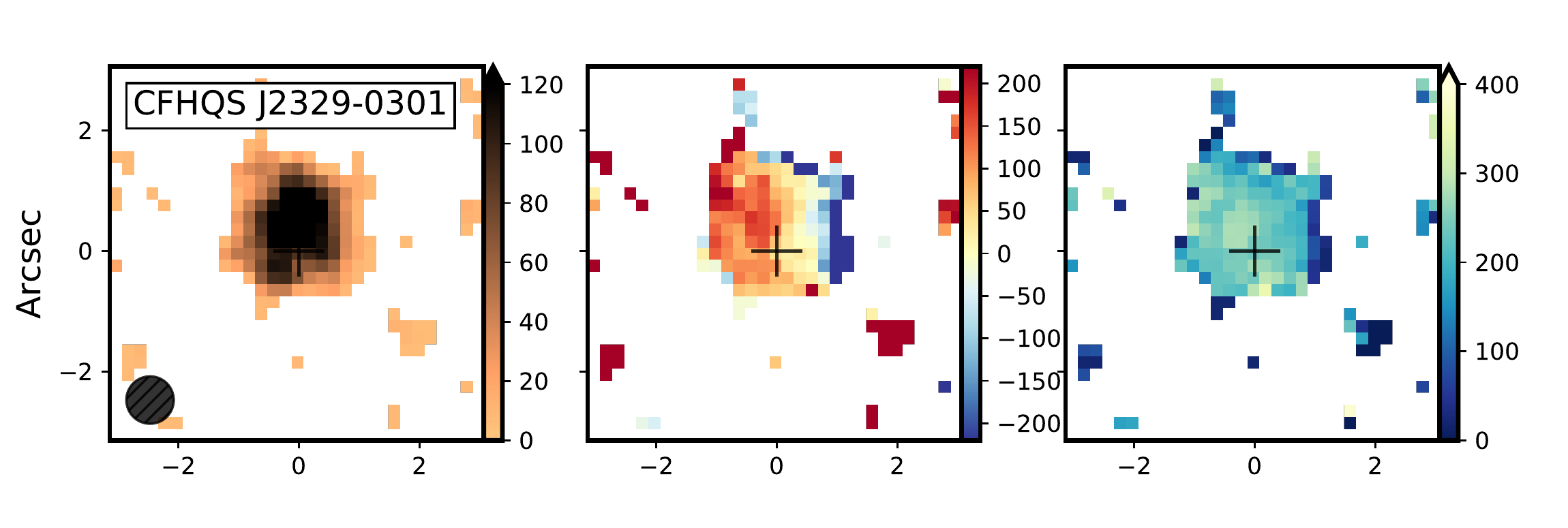}
	\includegraphics[width=0.9\textwidth]{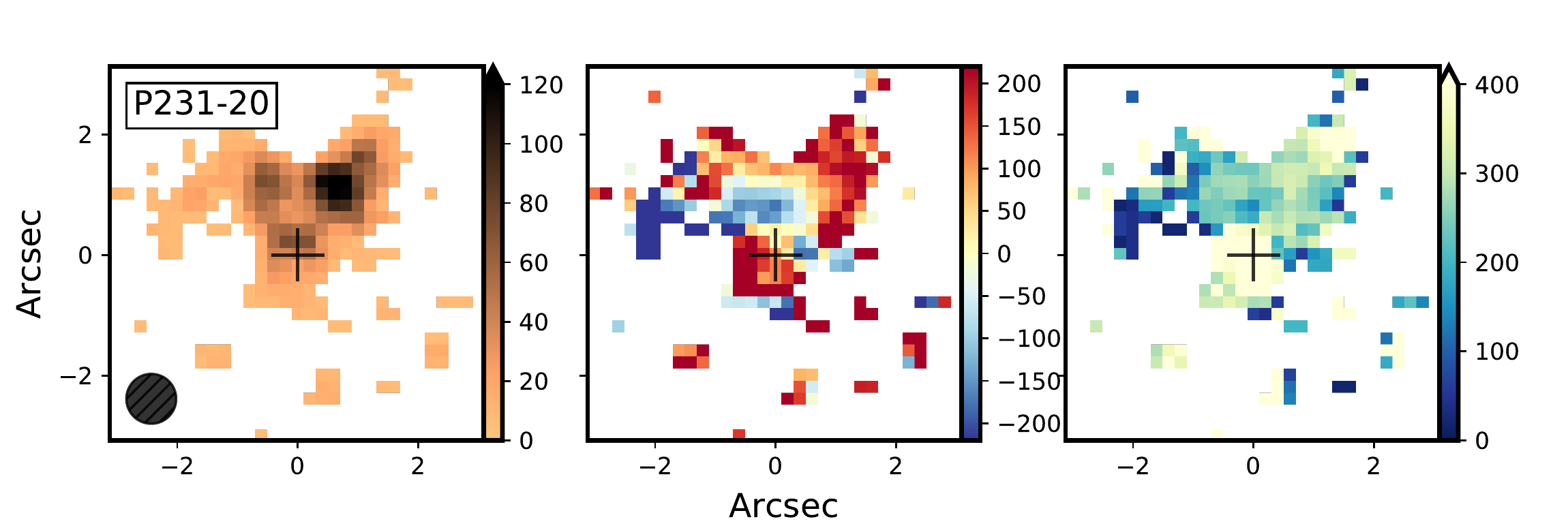}
    \caption{Ly$\alpha$ moment maps for the PSF-subtracted halos around each of the quasars. In the first column we show moment zero (flux images) containing all volumetric pixels with a S/N $\geq 2$, that make up the rest of the kinematic analysis. In the central column we show the first moment, i.e. the velocity of each spatial pixel relative to the peak of the halo Ly$\alpha$ emission, and in the right-hand column we show the second moment, or velocity dispersion in each spatial pixel. Details of the routine are given in Section \ref{Sect:ResKinematics}.}
	\label{Fig:MMM}
\end{figure*}

\begin{figure*}
    \centering
	\includegraphics[width=\columnwidth]{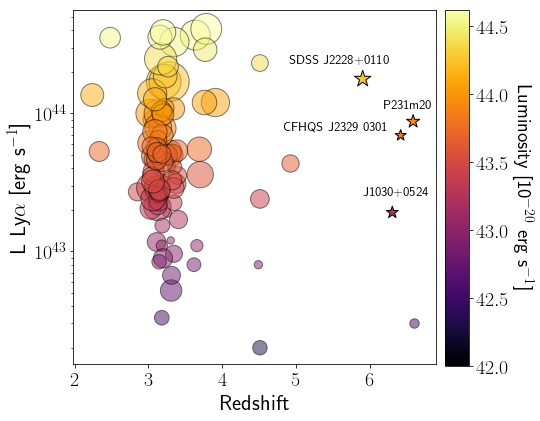}
	\includegraphics[width=\columnwidth]{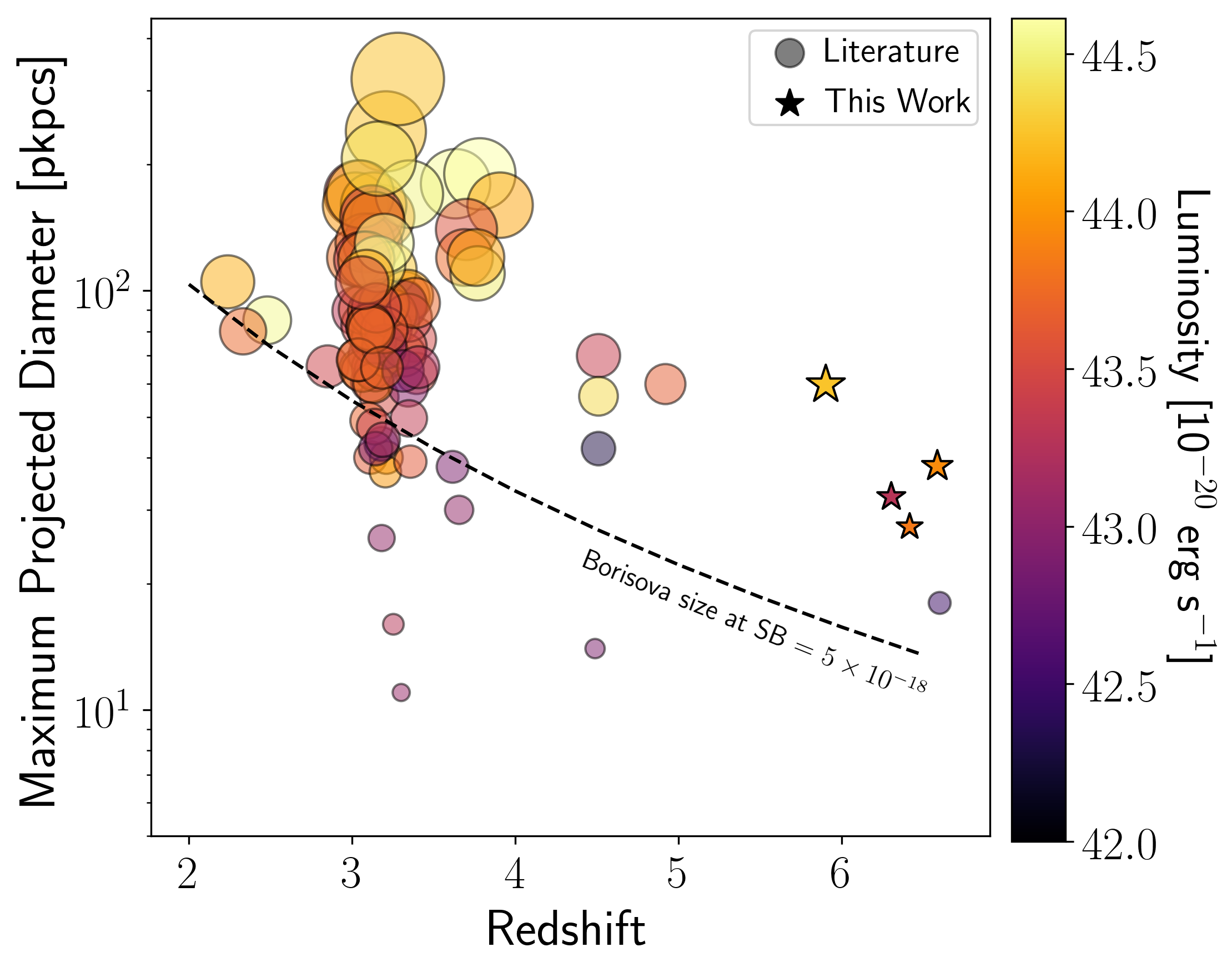}
	
	\caption{Total Ly$\alpha$ luminosities and sizes of Ly$\alpha$ halos surrounding QSOs plotted as a function of their redshifts. In each panel we show a compilation of literature sources from \cite{ArrigoniBattaia2018} and \cite{Borisova2016} (circular symbols) across the redshift range $2 < z < 5$. At redshift $z > 6$ we show the data point from \cite{Farina2017}, and the measurements presented in this work (star symbols). In each panel we plot halo redshift on the abscissa, and a commonly measured halo property on the ordinate; in the left-hand panel, measured total halo luminosities, and in the right-hand panel the measured maximum projected diameters in proper kiloparsecs, overlaid with a dashed line to show the apparent size of a typical \cite{Borisova2016} halo with redshift at our sensitivity limit. Size and luminosity information is also encoded in both panels as the size of the plotted points, and their colours, respectively.}
    \label{Fig: Halo properties with z}
\end{figure*} 

\subsubsection{Spatially resolved kinematics}
\label{Sect:ResKinematics}

In contrast to the integrated values, spatially-resolved kinematic maps give us an insight on the internal kinematics of the halos (as opposed to the bulk motion relative to the quasar discussed above in Section \ref{Sect:VelOff}). 

We keep the wavelength layers chosen in Section \ref{Sect:Halos}, smooth the datacube in the two spatial directions with a Gaussian kernel of $\sigma = 1.0$ pixel, and threshold the individual voxels (volumetric pixels) according to their S/N. In the first column of panels of Figure \ref{Fig:MMM} we show the zeroth-moment (flux images) that satisfy our S/N cut i.e. the voxels which make up the remainder of the kinematic analysis\footnote{Note that these images are distinct from the earlier fixed-width narrow-band images, as now each spatial pixel is made up of a different number of wavelength layers according to which met the S/N criterion}. Keeping all voxels with $S/N > 2.0$ we calculate the first- and second-moment maps. Again, much as in \cite{Borisova2016} and \cite{ArrigoniBattaia2018}, we do not assume any particular form of the Ly$\alpha$ line (i.e. we do not attempt to fit a Gaussian profile), and instead compute the non-parametric flux-weighted moments.

\begin{figure}
    \centering
	\includegraphics[width=\columnwidth]{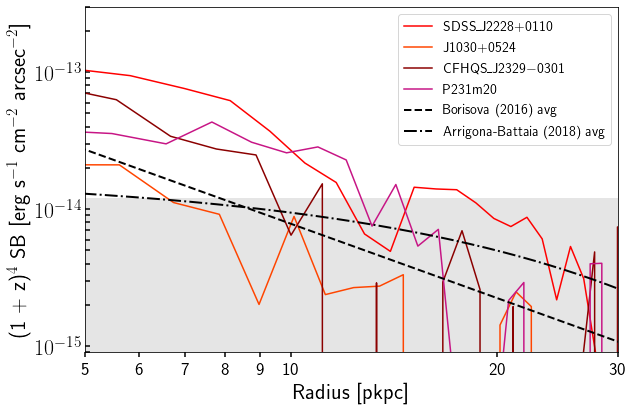}
	\caption{Azimuthally-averaged, surface brightness profiles in proper kiloparsecs corrected for surface-brightness-dimming (i.e. expressed in units of observed SB $\times (1 + z)^4$) for each of our objects. The shaded region indicates our surface brightness limit. We overplot the average profiles of \cite{Borisova2016} and \cite{ArrigoniBattaia2018}.}
    \label{Fig: SB profs}
\end{figure} 

\begin{figure*}
    \centering
	\includegraphics[width=\columnwidth]{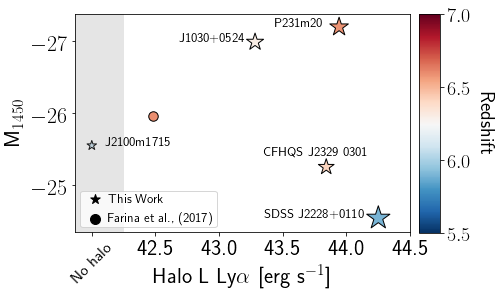}
	\includegraphics[width=\columnwidth]{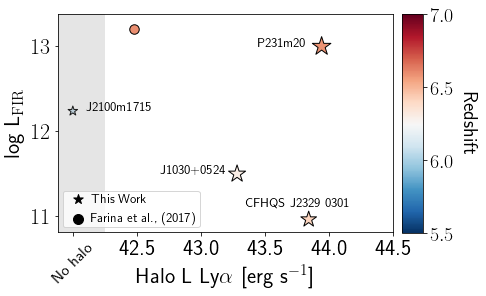}

	\caption{Observed quasar properties plotted against Ly$\alpha$ halo luminosities for all sources presented in this work, and the halo in \cite{Farina2017}. In the left-hand panel we show the absolute magnitude at $1450$ \AA, and in the right-hand panel the far-infrared luminosity. Points in each panel are colour-coded by redshift.}
    \label{Fig: QSO properties vs halos}
\end{figure*}

In the central column we show the first moment, i.e. the flux-weighted velocity of the halos relative to the peak of the Ly$\alpha$ emission from the halo, to provide information on the internal velocity structure of the halos. In these maps we can search for any sign of ordered motion -- although the Ly$\alpha$ line is not necessarily the best tracer of gas motion due to complex radiative transfer issues, we can use the velocity offset of the emission to look for any obvious signatures.

 The four halos present a complex picture, which is difficult to interpret. Each halo shows a velocity gradient across its spatial extent, indicating some movement in the gas, with total velocity ranges of $\Delta v \sim 400$ kms$^{-1}$ (i.e. $\pm 200$ kms$^{-1}$ relative to the peak of the Ly$\alpha$ from the halo).

In the case of \Roche, the only known radio-loud quasar in the sample, the gradient occurs in an East-West direction (i.e. {\emph{not}} varying with distance from the quasar) and with a slight warp. \Italian\ on the other hand displays a tentative velocity gradient with distance from the quasar, gas at the spatial position of the quasar is redshifted by $\sim 200$ kms$^{-1}$, but by the outer extents of the halo has shifted to  $\sim -200$ kms$^{-1}$. \CFHQS\ again shows some signs of ordered motion, with a velocity gradient and warp. Finally, \Ffav, the highest redshift object in our sample, displays some complex patterns. The quasar resides towards the South of the halo, with a patch of redshifted emission below it. Northwards of the quasar the emission is however blueshifted, and further North still, the offset-blobs of emission are again redshifted ($\sim 200$ kms$^{-1}$) relative to the quasar.

In the right-hand column of images we show the second moment, or velocity dispersion, $\sigma$, within the halos\footnote{Note that here we are displaying $\sigma$, and not the Gaussian-equivalent FWHM value in each pixel (frequently, $2.35$ times $\sigma$ is used in the literature.)}. Each halo shows a very broad Ly$\alpha$ line. The radio-loud quasar \Roche, does display the highest velocity dispersion, as one might expect following literature results such as \cite{Borisova2016} and \cite{Ginolfi2018}. The next two highest-redshift objects in our sample \Italian\ and \CFHQS\ both show a lower velocity dispersion. \Italian\ appears to show uniform values across its full extent, whereas for \CFHQS\ larger velocity dispersions are seen closer to the position of the quasar. Finally \Ffav\ again displays complex information. Here, the quasar is certainly centred in the  region of highest velocity dispersion, but there is a secondary peak in the image at the position of the large blob of emission in the North-Westerly direction.

\section{Discussion}
\label{Sect: Discussion}
\subsection{Comparison to Lower Redshift Samples}
\label{Sect:low-z}

Many results in the literature have noted that Ly$\alpha$ SB profiles, halo Ly$\alpha$ luminosities, and maximum projected sizes show smaller values at z$\sim$6 than at lower redshift. In Figure \ref{Fig: Halo properties with z} we present our measured Ly$\alpha$ halo luminosities and maximum projected sizes in the context of a compilation of literature sources. We include in the plot Ly$\alpha$ halos around all QSOs (both radio-loud and radio-quiet) from \cite{Borisova2016} and \cite{ArrigoniBattaia2018} in addition to the compilation presented in the PhD thesis of Arrigoni-Battaia\footnote{https://www.imprs-hd.mpg.de/49473/thesis\_Arrigoni.pdf} across the redshift range $2 < z < 5$. At redshift $z > 6$ we show the measurements presented in this work, and the data point from \cite{Farina2017}. 

\subsubsection{Evidence for Luminosity Evolution?}
Our measurements show little evolution of the luminosity of Ly$\alpha$ halos with redshift. This is in contrast to previous measurements at $z \ge 6$. Our re-analysis of \Roche\ in the MUSE datacube places its luminosity an order of magnitude higher, and likewise, \Italian, which previously was not reported to display a Ly$\alpha$ halo, now appears with a small halo of moderate luminosity, that is an order of magnitude more luminous than \cite{Farina2017}. This stresses the need for larger samples of Ly$\alpha$ halos around QSOs at $z \sim 6$.

\subsubsection{Evidence for Size Evolution?}
 In the right-hand side of Figure \ref{Fig: Halo properties with z} we show the maximum-projected diameters of our sample and those in the literature\footnote{Note: we now discuss {\emph{diameters}} of the halos by taking $2 \times$ the maximum radial projections measured in this work. This is for consistency with literature results, despite the fact that many reported halos, and all those presented in this work, are asymmetric.}. The plot appears qualitatively much as figure 6 of \citealt{Ginolfi2018}, who also note the apparent trend of increasing maximum projected size towards lower redshift. This size evolution is also corroborated by \cite{Momose2018}, who argue that the Ly$\alpha$ halo sizes scale with the size of evolving dark matter halos. Although we confirm with our data that the proper sizes are smaller at higher redshift, we do not propose any specific scaling in this work.
 
 We next compare our measurements to the average surface brightness profile measured in \cite{Borisova2016} at $\langle$$z = 3.2$$\rangle$. If we assume that this halo is typical of quasar (Ly$\alpha$) halos across cosmic time, we can predict its appearance at redshift z in observations reaching our average surface brightness limit (for the sources measured in this paper, the sizes are measured on a background of a ($ \sim$ $3-7$)$\times$ $10^{-18}$ \SBunits\, see Table \ref{tab:tot_fluxes}). For instance, we can see from Figure \ref{Fig: Halo properties with z}, that at $z\sim6$ the value of the \cite{Borisova2016} halo's maximum projected diameter would be measured as $\le 15$ pkpc in observations reaching a surface brightness level of SB $\sim$ \num{5.0E-18} \SBunits.  This means that each of the Ly$\alpha$ halos presented in this work appears larger than a typical \cite{Borisova2016} halo would appear placed at $z\sim6$. 
 
\subsubsection{Evidence for Surface Brightness Profile Evolution?}

In Figure \ref{Fig: SB profs} we present the azimuthally-averaged surface brightness profiles of each of the quasars presented here. We show the surface-brightness-dimming corrected profiles (i.e. SB$_{\mathrm{obs}}$ $\times$ $(1 + z)^4$) on a scale of proper kiloparsecs (pkpc). The grey shaded area shows our surface brightness sensitivity. A comparison to \cite{Borisova2016} and \cite{ArrigoniBattaia2018} shows that we would barely have detected their average profiles in our observations.  

\begin{figure}
    \centering
	\includegraphics[width=0.95\columnwidth]{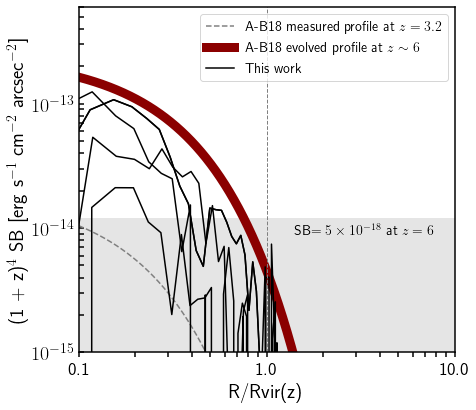}
	\caption{Surface brightness profiles of our Ly$\alpha$ halos normalised to the virial radius for a dark matter halo of mass log M$_{\mathrm{DM halo}} = 11.5$ M$_\odot$ (see text). We show the Ly$\alpha$ halos measured in this work in thick black lines, and the measured profile of \cite{ArrigoniBattaia2018} at $\langle z=3.2 \rangle$ in a grey dotted line. The thick red band shows the appearance of the \cite{ArrigoniBattaia2018} profile assuming that the SB profile evolution found by \cite{ArrigoniBattaia2018} at $z=2-3$ holds out to $z\sim6$. The surface brightness profiles are surface-brightness-dimming-corrected (i.e. $[1 + z] \times$ SB) and measured radii are in proper units, normalised to a virial radius calculated according to {\mbox{R$_{vir}$ $=$ $[3 M_{\mathrm{DM halo}} / (800 \pi \rho_{crit}(z))]^{\frac{1}{3}}$}}.}
    \label{Fig: R_vir}
\end{figure}

\subsection{Physical Origin of Ly$\alpha$ emission}
\label{Sect:origin lya-em}

In order to determine the energy source and spatial origin of the Ly$\alpha$ emission observed here as a halo, we need additional line diagnostics from the same gas. With Ly$\alpha$ alone, we can only use the velocity width and spatial extent on-sky to speculate on the physical processes at work. In addition, we can use previous observations of the QSOs to look for any correlation between the observed properties of the QSOs and the Ly$\alpha$ halos.

In Figure \ref{Fig: QSO properties vs halos} we plot for each source its halo luminosity against an observable commonly used to characterise the quasars themselves. In the left-hand panel we show Ly$\alpha$ halo luminosity against M$_{\mathrm{1450}}$, which can also be used as a proxy of the black hole mass. The scatter of the measurements in this plot does not imply any direct link-with-/influence-by the black hole on the Ly$\alpha$ halo. Indeed, \Comp\ which displays no halo above the limit of our observations is firmly towards the middle of the range of M$_{\mathrm{1450}}$ values, as is the data point from \cite{Farina2017} which displays only a faint halo\footnote{Note that although black-hole mass estimates exist for several of the objects examined here, these rely on additional assumptions, are derived via different methods for different QSOs depending on the information available, therefore we choose to examine only the M$_{\mathrm{1450}}$ proxy for each object.}, however our dynamic range is indeed limited, and we present only \nohalos\ halos. In the right-hand panel we show the halo luminosity again against the log of the far-infrared luminosity, tracing obscured star-formation. Note that unfortunately no information is available for \Roche. Again the points show no obvious correlation between star formation activity in the host galaxy and the powering of the Ly$\alpha$ halo. These results point towards a scenario in which the Ly$\alpha$ halo emission may be more closely related to properties of diffuse gas in the CGM/IGM than of the black hole or host galaxies' stellar populations.
 
 \subsection{Ly$\alpha$ Emission within R$_{\mathrm{vir}}$?}
\label{Sect: R vir}
 A particularly interesting question is whether we are observing pristine gas in the IGM falling onto the QSOs, or gas that is part of the objects' CGM i.e. within the virial radii. To make any statement about the extent of the emitting gas with respect to each object's virial radius, we need to make some assumption about the hosting M$_{\mathrm{DM halo}}$. Evidence suggests that black holes are already $10^9$ M$_{\mathrm{\odot}}$ at $z\sim6$ (\citealt{Mazzucchelli2017}), and that gas masses are of order $10^{10}$ M$_{\mathrm{\odot}}$ (\citealt{Venemans2017a}, \citealt{Venemans2018}), and the sum of these components will be {\mbox{$\ll$ M$_{\mathrm{DM halo}}$}}. As such, we assume a conservative lower limit of log$_{\mathrm{10}}$ M$_{\mathrm{DM halo}} = 11.5$ to compute R$_{\mathrm{vir}}$ according to Equation 1 (as in \citealt{ArrigoniBattaia2018}):

\begin{equation}
    R_{vir} = [3 M_{\mathrm{DM halo}} / (800 \pi \rho_{crit}(z))]^{\frac{1}{3}}
\end{equation}

\noindent where $\rho_{crit}(z)$ is the critical density at redshift z. For log$_{\mathrm{10}}$ M$_{\mathrm{DM halo}} = 11.5$ at $z=6$ this gives a virial radius of R$_{{\rm{vir}}} \approx 30$ pkpc. Each of the Ly$\alpha$ halos presented in this work display a maximum radial extent $\lesssim 30$ pkpc, meaning that, unless the hosting dark matter halo mass {\mbox{$\ll$ log$_{\mathrm{10}}$ M$_{\mathrm{DM halo}} = 11.5$}} (which we consider unlikely) we are observing emission from circum-galactic gas {\emph{inside}} the virial radius i.e. not the IGM. To investigate the nature of this gas further (pristine? first in-fall? outflows?) requires additional diagnostics which are not currently available.

\subsection{Ly$\alpha$ Halos Consistent with Evolution Seen at $z=2-3$?}
\label{Sect: FAB evo}

\cite{ArrigoniBattaia2018} derive an empirical model of Ly$\alpha$ halo evolution from $z=3.2$ to $z=2.25$. They find that the normalisation of their surface brightness profiles decreases by approximately an order of magnitude over this redshift range. Assuming that this trend holds out to $z=6$, we estimate the resulting profile and compare to our observations in Figure \ref{Fig: R_vir}, assuming the virial radius estimates from Section \ref{Sect: R vir}.  Our SB profiles fall below such an empirically predicted evolution. This means that the evolution between the average SB profile at $z\sim3$ and $z\sim6$ is not as strong as that seen between $z\sim3$ and $z\sim2$ in \cite{ArrigoniBattaia2018}.

\section{Conclusions}

We present MUSE (archival) data of \noobjs\ $z\sim6$ QSOs. After PSF-subtraction we search for extended Ly$\alpha$ emission in the vicinity of each quasar. For \nohalos\ out of \noobjs\ QSOs we detect extended, diffuse Ly$\alpha$ emission, directly probing the CGM at $z\sim6$. 

Our findings are summarised below:

\begin{enumerate}

\item The four
Ly$\alpha$ halos presented here are diverse in morphology and size, they each
display spatial asymmetry, and none are centred on the position of the
quasar.

\item None of the halos are significantly offset in velocity from the systemic redshift of the quasars ($\Delta$ v $< 200$
kms$^{-1}$).

\item Each halo shows a broad
Ly$\alpha$ line, with a velocity width of order $1000$ kms$^{-1}$. 

\item Total Ly$\alpha$ luminosities range between
$\sim$\,{\num{2E43}}\, erg s$^{-1}$ and $\sim$\,{\num{2E44}}\, erg s$^{-1}$, reaching 
maximum radial extents of $13 - 30$ pkpc from the quasar positions.

\item We find larger sizes and higher Ly$\alpha$ luminosities than previous literature results at this redshift (that generally did not employ IFU data). This alters the perception that Ly$\alpha$ halos are less luminous at higher redshift. 

\item We see no correlation between QSO properties and the characteristics of the Ly$\alpha$ halo, and thus infer no evidence that black holes or the stellar populations of host galaxies are the primary driver of the Ly$\alpha$ halo emission at $z\sim 6$.

\item The Ly$\alpha$ emission observed here is located within the virial radius of our targets, assuming a conservative log$_{\rm{10}}$ M$_{{\rm{DMhalo}}} = 11.5$. 

\item The redshift evolution of the SB profiles between $z\sim6$ and $z\sim3$ appears to be less pronounced than seen between $z\sim3$ and $z\sim2$ in \cite{ArrigoniBattaia2018}. 


\end{enumerate}

Overall, our results are consistent with a picture in which the physical properties of the CGM evolve with cosmic time, manifesting as an observed evolution of Ly$\alpha$ halo properties. Before this scenario can be confirmed or clarified, a larger sample of QSO observations at $z\sim6$ is called for, and observations of more diagnostic lines (e.g. with the upcoming JWST mission). This will help to elucidate the processes governing the growth of the first galaxies and black holes.

\acknowledgments

The authors wish to thank the anonymous referee for insightful suggestions that have improved the quality of this work. ABD, MN and FW acknowledge funding through the ERC grant ``Cosmic Gas". ABD acknowledges the MUSE Python Data Analysis Facility, ``MPDAF", developed at Centre de Recherche Astrophysique de Lyon. This work is based on observations collected at the European Southern Observatory under ESO programmes: 095.B-0419(A), 297.A-5054(A), 095.A-0714(A), 60.A-9321(A), and 099.A-0682(A).

\appendix

\section{Choosing the Spectral window}

\begin{figure*}
\centering
    \includegraphics[width=0.65\textwidth]{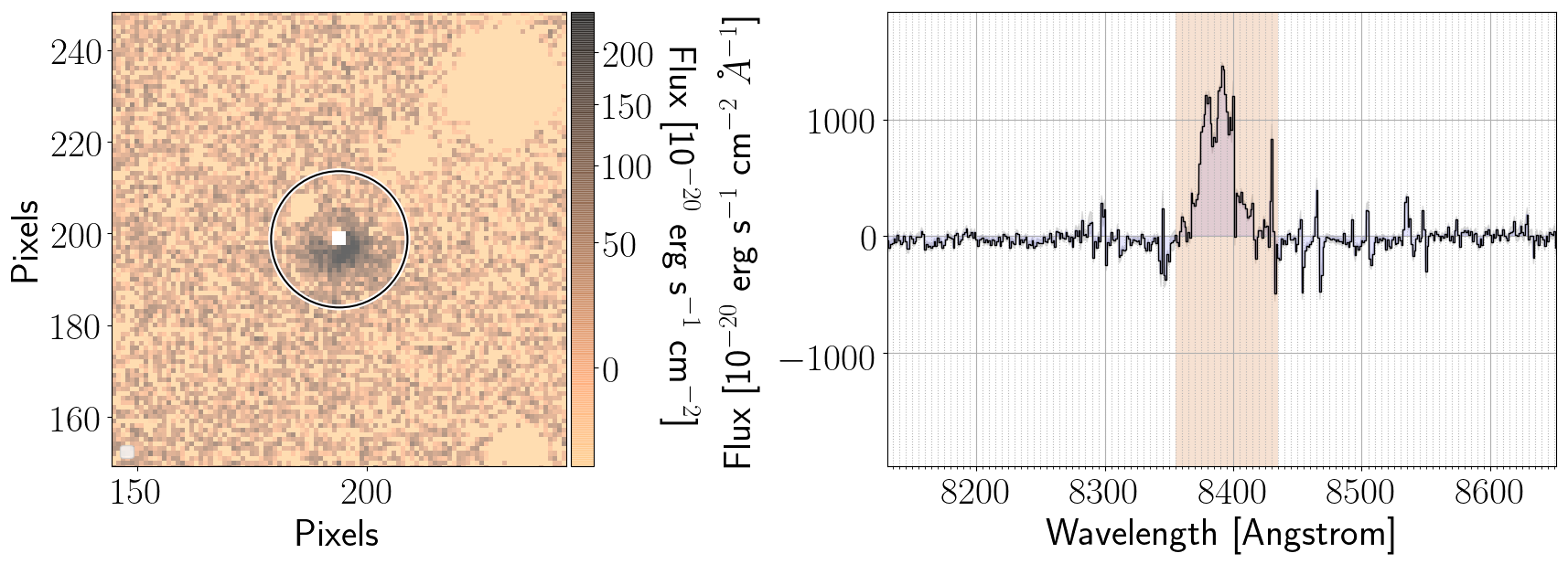}
    \includegraphics[width=0.65\textwidth]{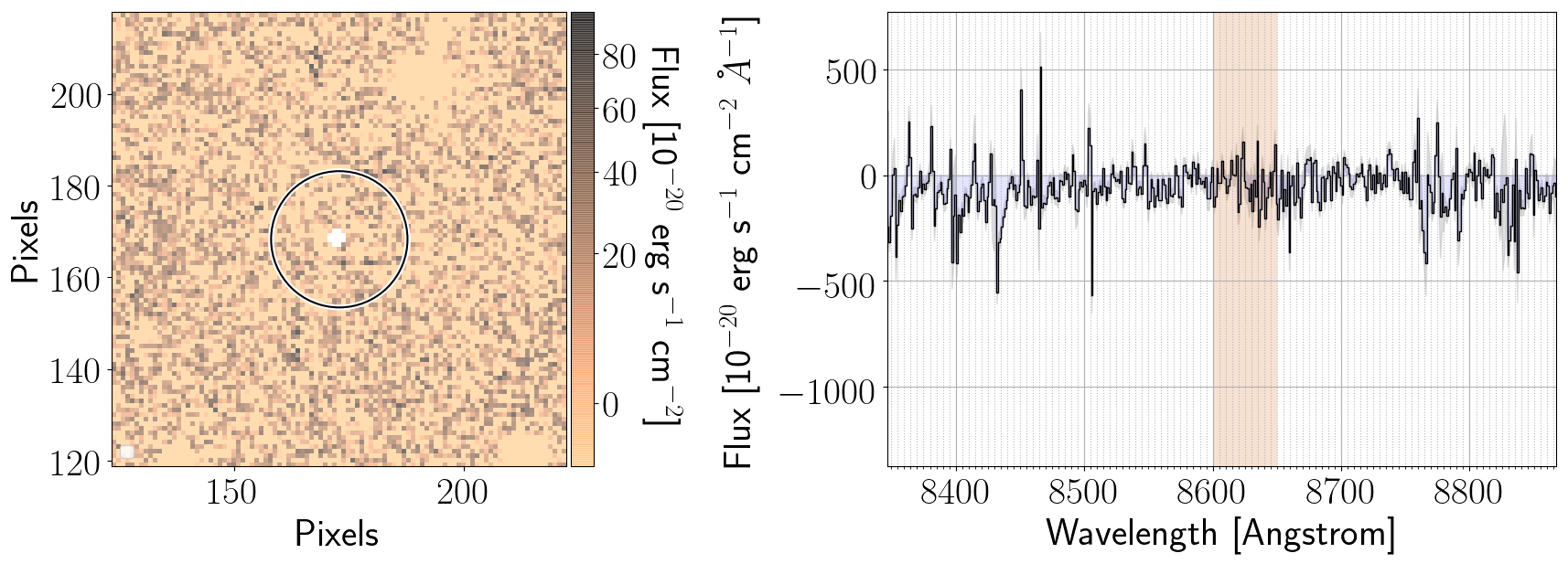}
    \includegraphics[width=0.65\textwidth]{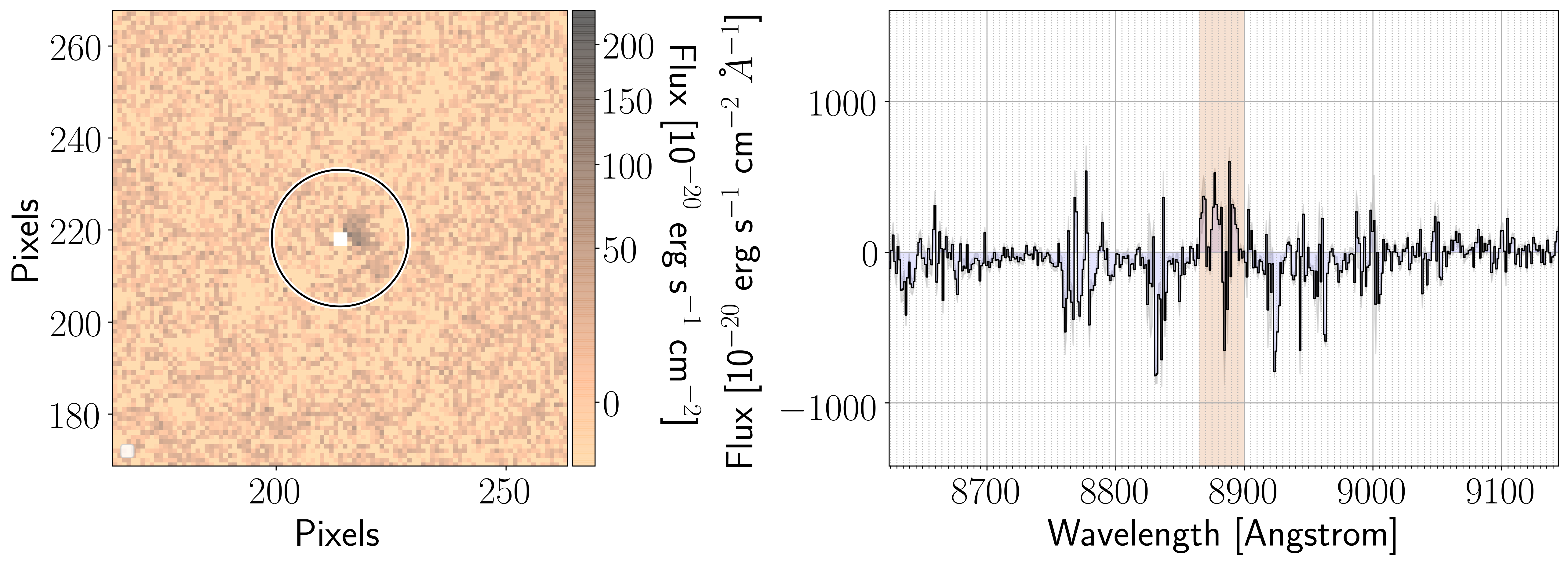}
    \includegraphics[width=0.65\textwidth]{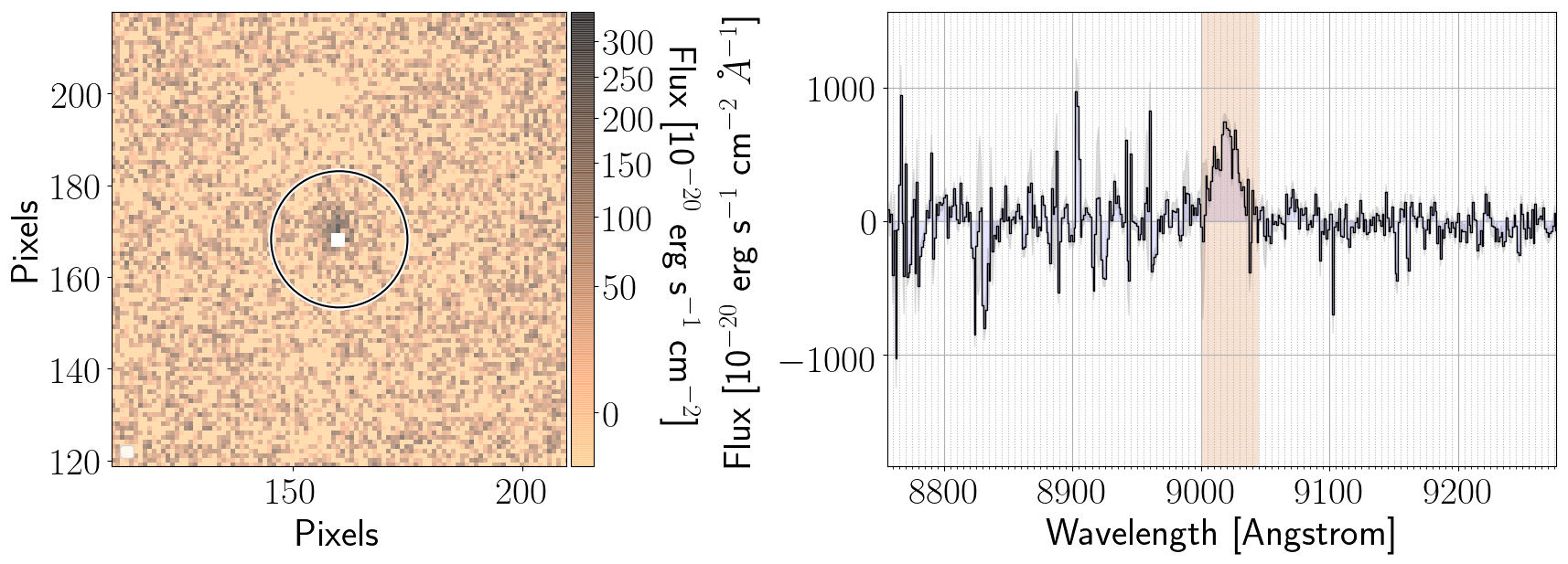}
	\includegraphics[width=0.65\textwidth]{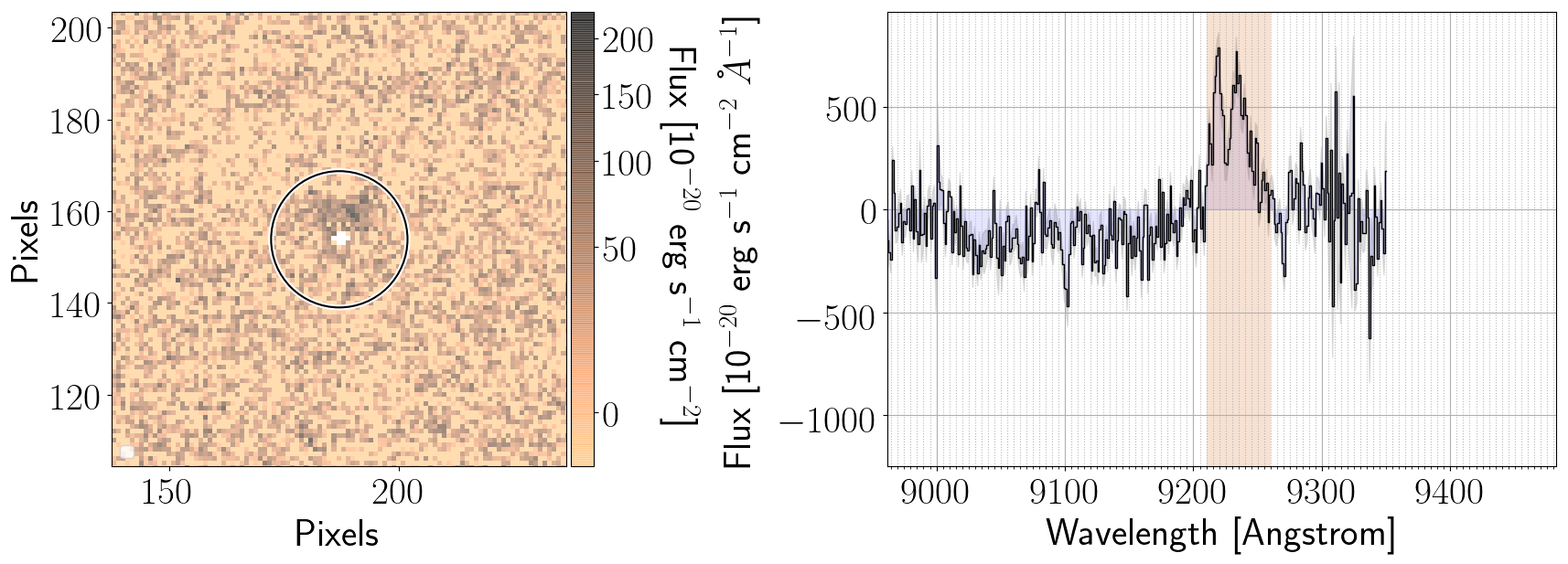}
    \caption{Spectral windows over which we choose to perform photometry of each PSF-subtracted halo. The left-hand panels show narrow-band images, with an aperture of $6$ arcseconds overlaid, and the pixels excluded from the Ly$\alpha$ halo analysis due to complex PSF-subtraction residuals are masked. In the right-hand column of panels we show the spectra extracted from within the $6$ arcsecond diameter aperture from which the wavelength layers to make up the fixed-width narrow-band image were chosen by eye. The wavelength range chosen is highlighted in orange on the spectrum. These wavelength ranges are then propagated to the total-flux measurements (Section \ref{Sect:Fluxes}, and Figure \ref{Fig:tot_flux}), and the surface brightness images from which we analyse the `bulk properties' of each halo (Section \ref{Sect:UniSB}, and Figure \ref{Fig:2_blobs}).}
	\label{appendix}
\end{figure*}

\section{White-Light Images}

\begin{figure*}
\centering
    \includegraphics[width=0.4\textwidth]{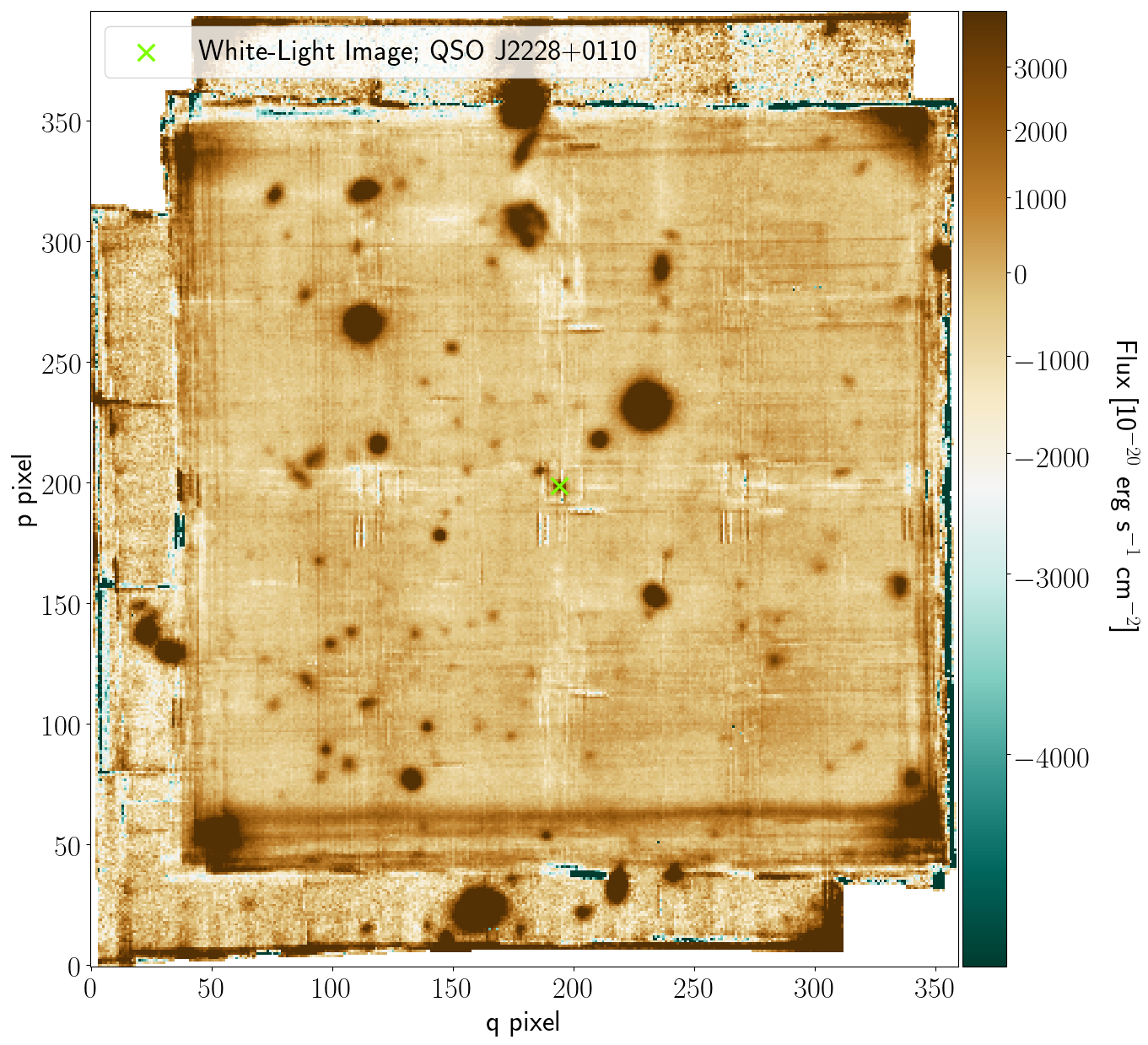}
    \includegraphics[width=0.4\textwidth]{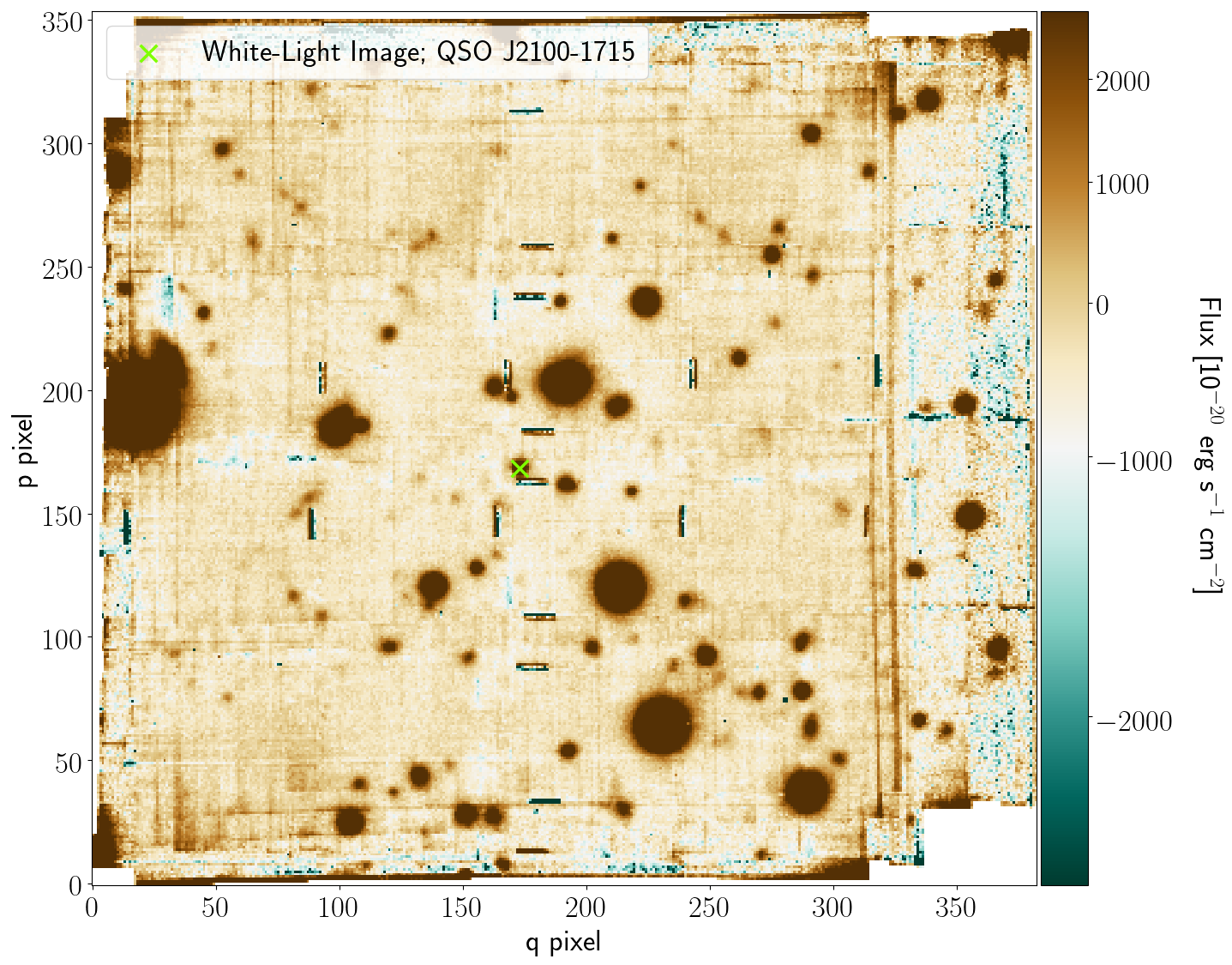}
    \includegraphics[width=0.4\textwidth]{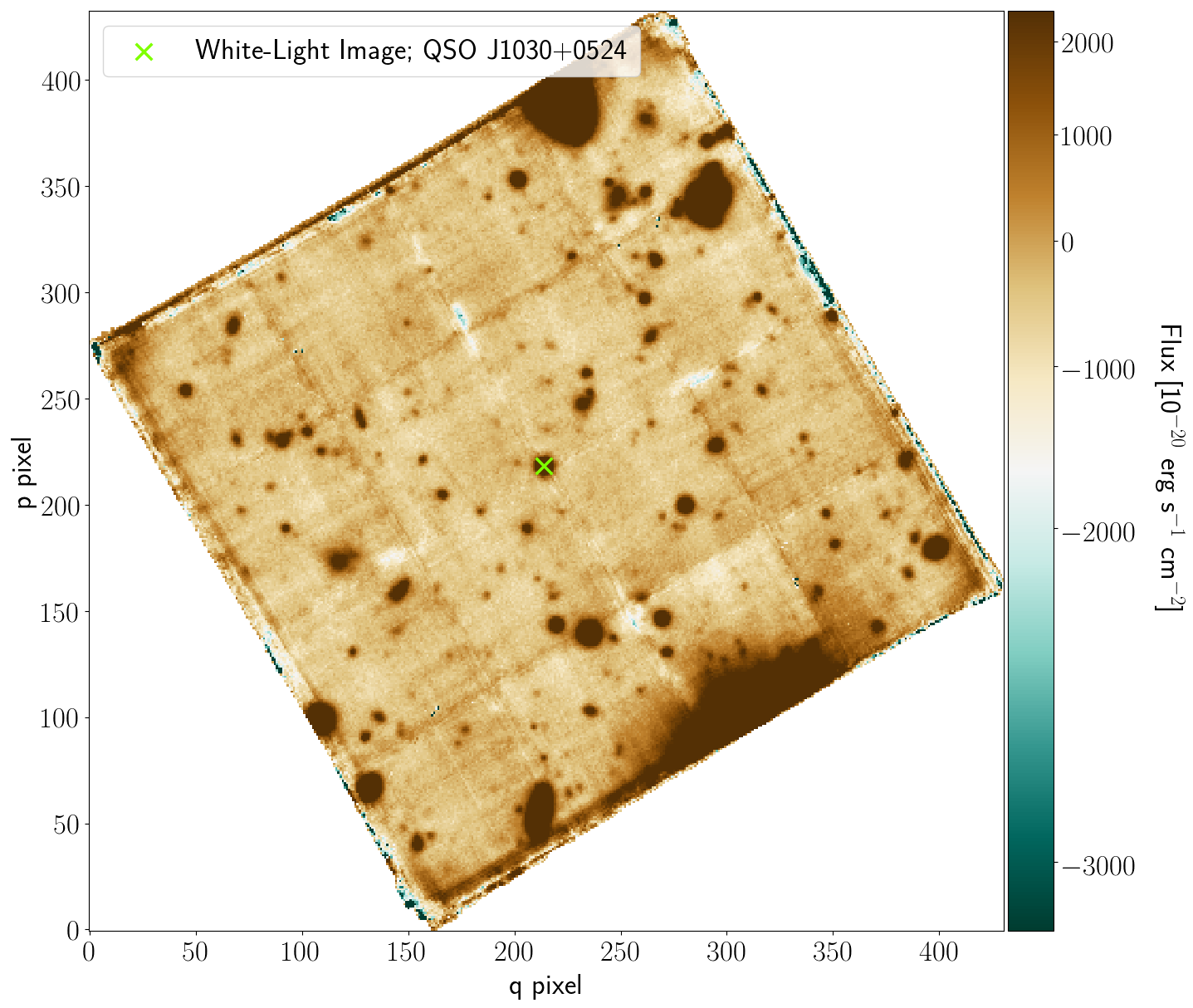}
    \includegraphics[width=0.4\textwidth]{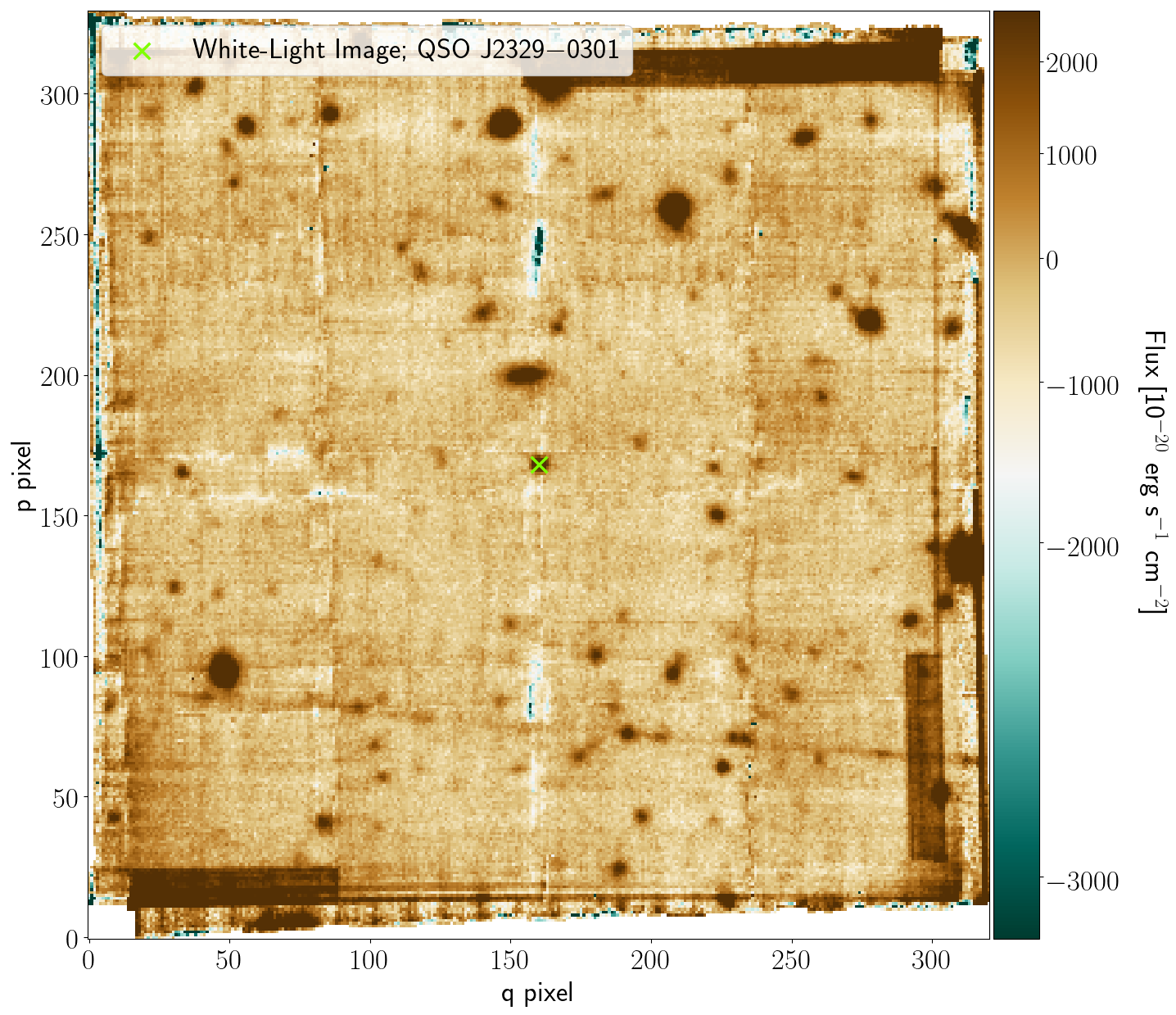}
    \includegraphics[width=0.4\textwidth]{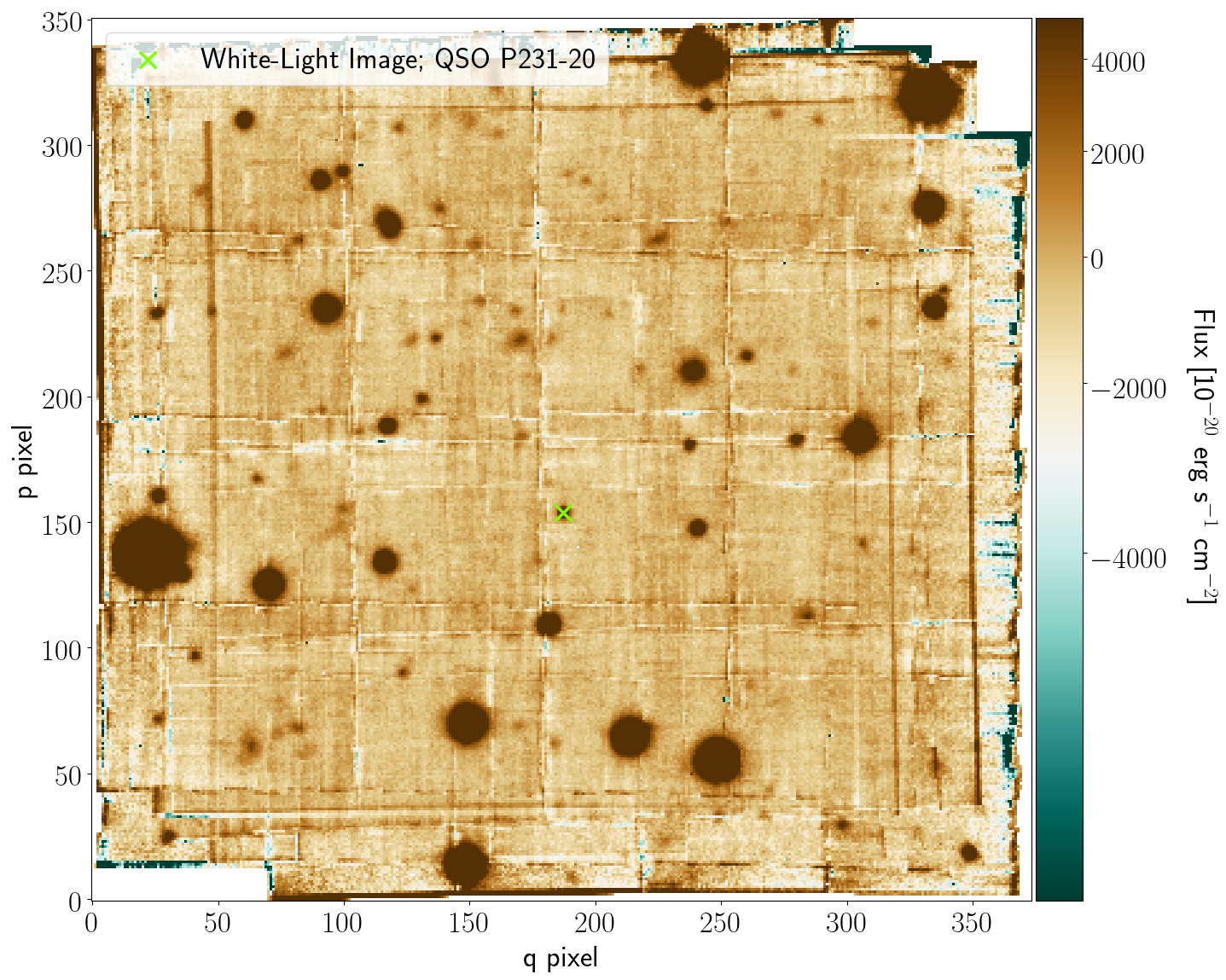}
    \caption{White-light images of each ZAPPED MUSE datacube used in this analysis. In each cutout the quasar is marked by a green cross. The images demonstrate that our measurements are not affected by very bright continuum objects in close proximity to any of the quasars.}
	\label{appendix}
\end{figure*}


\end{document}